\documentclass[12pt]{article}
\usepackage{amsmath}
\usepackage{graphicx}
\usepackage{enumerate}
\usepackage{natbib}
\usepackage{url} 

\newcommand{\blind}{1}

\usepackage{times}

\usepackage{natbib}

\usepackage{latexsym}
\usepackage{bm}

\usepackage{setspace}
\usepackage{lineno}

\usepackage{amssymb}
\usepackage{color,pgf}
\usepackage{multimedia}

\usepackage{arydshln}

\begin{document}

\newcommand{\dif}[2]{\frac{{\rm d} #1}{{\rm d} #2}}
\newcommand{\ddif}[3]{\frac{{\rm d}^2 #1}{{\rm d} #2 {\rm d} #3}}
\newcommand{\ildif}[2]{{\rm d} #1/{{\rm d} #2 }}
\newcommand{\ilpdif}[2]{\partial #1/{\partial #2 }}
\newcommand{\pdif}[2]{\frac{\partial #1}{\partial #2}}
\newcommand{\pddif}[3]{\frac{\partial^2 #1}{\partial #2 \partial #3}}
\newcommand{\ilpddif}[3]{\partial^2 #1/{\partial #2 \partial #3}}
\newcommand{\comb}[2]{\left (\begin{array}{c}{#1}\\{#2}\end{array}\right )}
\newcommand{\perm}[2]{^{#1}{\rm P}_{#2}}
\newcommand{\gfrac}[2]{\mbox{$ { \textstyle{ \frac{#1}{#2} }\displaystyle}$}}
\newcommand{\defn}{\begin{quote}{\bf Definition. }}
\newcommand{\edefn}{\end{quote}}
\newcommand{\thm}{\begin{theorem}}
\newcommand{\ethm}{\end{theorem}}
\newcommand{\its}{^{\sf -T}} 
\newcommand{\fv}{\hat{\bm{\mu}}}
\newcommand{\X}{{\bf X}}
\newcommand{\Xt}{\X\ts}
\newcommand{\y}{{\bf y}}
\newcommand{\A}{{\bf A}}
\newcommand{\bp}{{\bm \beta}}
\newcommand{\bg}{{\bm \gamma}}
\newcommand{\bt}{{\bm \theta}}

\newcommand{\bmat}[1]{\left [ \begin{array}{#1}}
\newcommand{\emat}{\end{array}\right ]}
\newcommand{\E}{\mathbb{E}}
\newcommand{\beq}{\begin{equation}}
\newcommand{\eeq}{\end{equation}}
\newcommand{\Gi}{{\bf G}^{-1}}
\newcommand{\B}{{\bf B}}
\newcommand{\Bt}{{\bf B}\ts}
\newcommand{\D}{{\bf D}}
\newcommand{\V}{{\bf V}}
\newcommand{\p}{{\bf P}}
\newcommand{\K}{{\bf K}}
\newcommand{\Uo}{{\bf U}_1}
\newcommand{\Tk}{{\bf T}_k}
\newcommand{\Tm}{{\bf T}_m}
\newcommand{\Tkm}{{\bf T}_{km}}

\newcommand{\mc}{0}  

\newcommand{\ts}{^{\rm T}}
\newcommand{\st}{^{\star}}
\newcommand{\stt}{^{\star \rm T}}
\newcommand\Nicole[1]{{\color{blue} #1}}

\newcommand{\eps}[3]
{{\begin{center}
 \rotatebox{#1}{\scalebox{#2}{\includegraphics{#3}}}
 \end{center}}
}
\newcommand{\tr}{{\rm tr}}

\def\spacingset#1{\renewcommand{\baselinestretch}%
{#1}\small\normalsize} \spacingset{1}


\if1\blind
{
\title{\bf Modelling tree survival for investigating climate change effects}
\author{Nicole Augustin,  
School of Mathematics, University of Edinburgh, UK\\
\hspace{.2cm}\\ 
Axel Albrecht ,  
Forest Research Institute Baden-W\"urttemberg, Freiburg, Germany\\
\hspace{.2cm}\\ 
Karim Anaya-Izquierdo, 
Department of Mathematical Sciences, University of Bath, UK\\
\hspace{.2cm}\\
Alice Davis, 
Mayden, The Old Dairy, Melcombe Road, UK\\
\hspace{.2cm}\\ 
Stefan Meining, 
Office of Environmental Monitoring, Freiburg, Germany\\\hspace{.2cm}\\ 
 Heike Puhlmann, 
Forest Research Institute Baden-W\"urttemberg,  Freiburg, Germany\\
\hspace{.2cm}\\ 
and Simon Wood, 
School of Mathematics, University of Edinburgh, UK}

  \maketitle
} \fi

\if0\blind
{
  \bigskip
  \bigskip
  \bigskip
  \begin{center}
 {\LARGE\bf Modelling tree survival for investigating climate change effects}
\end{center}
  \medskip
} \fi

\bigskip
\begin{abstract}

Using German forest health monitoring data we investigate the main drivers leading to tree mortality and the association between defoliation and mortality; in particular (a) whether defoliation is a proxy for other covariates (climate, soil, water budget); (b) whether defoliation is a tree response that mitigates the effects of climate change and (c) whether there is a threshold of defoliation which could be used as an early warning sign for irreversible damage. Results show that environmental drivers leading to tree mortality differ by species, but some are always required in the model. The defoliation effect on mortality differs by species but it is always strong and monotonic. There is some evidence that a defoliation threshold exists for spruce, fir and beech. 

We model tree survival with a smooth additive Cox model allowing for random effects taking care of dependence between neighbouring trees and non-linear functions of spatial time varying and functional predictors on defoliation, climate, soil and hydrology characteristics. 
Due to the large sample size and large number of parameters, we use parallel computing combined with marginal discretization of covariates. We propose a 'boost forward penalise backward' scheme based on combining component-wise gradient boosting 
with integrated backward selection.
\end{abstract}

\noindent%
\vfill

\spacingset{1.48} 
\section{Introduction} 
\label{sec:intro}
Forest health is monitored in Europe by The International Co-operative Programme on Assessment and Monitoring of Air Pollution Effects on Forests (ICP Forests, \cite{Eich2016}) in cooperation with the European Union. Recently climate change has contributed to the decline in forest health and these data are increasingly being used to investigate the effects of climate change on forests in order to decide on forest management strategies for mitigation.  Forests in Germany have been badly affected and climate change now appears to be the major cause of defoliation \citep{eickenscheidt2018spatio,Augetal09}.

Here we focus on two main questions to investigate climate change effects on tree mortality. These are, what are the main drivers leading to tree mortality; and, what is the association between defoliation and mortality? Regarding the second question, we explore (a) whether defoliation is a proxy for other covariates (climate, soil, water budget); (b) is defoliation a tree response that mitigates the effects of climate change 
and (c) whether there is a threshold of defoliation which could be used as an early warning sign for irreversible damage. If this threshold is found, it has practical relevance for forest management. Trees with defoliation greater than this threshold value but not dead yet could be harvested in an anticipatory manner or stabilised.
We focus in our analysis on the main species: Norway spruce, Silver fir, Scots pine, oak (Pedonculate oak and Sessile oak) and common beech. To answer these questions we use extensive yearly data on tree mortality and crown defoliation, an indicator of tree health, from a monitoring survey carried out in Baden-W\"urttemberg, Germany since 1983, which includes a part of the ICP transnational grid. On a spatial grid, defoliation, mortality and other tree and site specific variables are recorded.  In some cases the grid locations are no longer observed which leads to censored data. Also recruitment of trees happens throughout when new grid points are added or trees are replaced. 

There are a number of statistical challenges with regard to modelling tree survival for investigating climate change effects. The large number of correlated predictor variables seriously impacts the complexity of model selection. Some of these variables are time varying and many variables have non-linear effects. For example, there is an optimal range of Julian day of budburst implying a non-linear effect of this variable. At the survey grid locations, several trees, each a maximum of 50m apart, are observed over time and we need to account for this short range spatial correlation. An appropriate model for this type of short range correlation would include a spatial frailty term, i.e. a random effect for grid location. With more than one thousand locations, this results in some very large models. 

Several approaches have been applied to modelling tree survival.  
\cite{neuner2015survival} used parametric Weibull accelerated failure time models to analyse the survival of Norway spruce and European beech in Baden-W\"{u}rttemberg, Bavaria and Rhineland-Palatinate. In this study the survival time was age of tree at death. \cite{nothdurft2013spatio} looked at age-dependent survival of Norway spruce, Silver fir, Scots pine and common beech in Baden-W\"{u}rttemberg using a Cox model with a frailty term, to account for correlation between individuals in survey data which overlaps with ours. 
\cite{li2015survival} and \cite{thapa2016modeling} model the survival of the Loblolly Pine in the Piedmont, Atlantic Coastal Plain  and Gulf Coastal Plain regions of the US, also using Cox frailty models with special attention to the parametric form of the spatial correlation \citep{li2015survival}.  In  \cite{thapa2016modeling} the survival time is time in years  since plot establishment. \cite{lee2011comparison} use a survival model for interval censored data in a joint model of tree growth and mortality.  \cite{ZhouHanson2018}  fit a Bayesian
semiparametric model to arbitrarily censored spatially referenced survival data. 

We model tree hazard at calendar time using a Cox model as a function of the predictor variables on climate, soil characteristics and water budget.
 This approach assumes the baseline hazard changes with calendar time and  means the baseline hazard can explain some temporal trends common to all trees driven by hidden explanatory variables which change with calendar time. In contrast, others 
 \citep{neuner2015survival, nothdurft2013spatio} use tree age as the survival time and in that case the baseline hazard is a function of age and can explain trends common to all trees changing with age, but this would not be so suited to investigating the relationship between mortality and environmental variables. \cite{maringer2021ninety} model tree mortality from experimental research plots in Baden-W\"urttemberg, with stand age as the survival time, using an accelerated failure time model with linear effects. 

We use a smooth additive Cox  model for the hazard, which allows for a spatial frailty term taking care of dependence between neighbouring trees and non-linear smooth functions of spatial, time varying and functional predictors modelling any spatial trend. For parameter estimation, including smoothness parameters, we use a penalised version of the partial likelihood of the Cox model. 

In order to account for time varying covariates we use a Poisson generalised additive mixed model for pseudodata to fit the Cox model \citep{whitehead1980fitting}. This is possible because the Poisson likelihood for the pseudodata  is identical to the partial likelihood up to a constant of proportionality. This approach also accounts automatically for the varying risk set size induced by the sampling protocol, causing left truncation, and this is similar to the adjustment for left truncation in epidemiological cohort studies 
\citep{Thiebaut2004, Pencina2007}. As outlined in \cite{Wood17}, Section 7.8.1 and 
\cite{bender2018generalized} the use of a Poisson generalised additive mixed model allows the estimation of time-varying effects in these smooth additive Cox models. The computational cost due to data expansion in the GAMM approach is not worse than in the traditional partial likelihood based approach as soon as time varying predictors and non-linear effects are involved. In addition using the GAMM approach gives access to the entire GAMM tool box offered by the mgcv package in R. This includes the use of REML  estimation of smoothing and variance parameters, reliable and generally applicable tests for semiparametric terms, locally adaptive smooths, and integrated backward selection.  

Altogether a total of 57720 trees are observed making the analysis computationally challenging  
due to the large number of parameters and large sample sizes once pseudodata are generated. To this end, we use the efficient model estimation methods of \cite{wood2017generalized} which combine parallel computing with marginal discretization of covariates.
 This  is implemented in the function $\tt bam$ in the R package {\tt mgcv} in R also facilitating feasible model selection.

To select from a large number of correlated environmental predictors with non-linear effects we propose a `boost forward penalise backward' model selection scheme based on combining component-wise gradient boosting \citep{schmid2008boosting,mayr2012generalized}  with integrated backward selection \citep{marra2011practical}.
In addition this new model selection approach also works in applications where the number of parameters is larger than the sample size.
\section{Data}
We model tree survival data from the Terrestrial Crown Condition Inventory (TCCI), a forest health monitoring survey which has been carried out yearly in the forests of Baden-W\"urttemberg since 1983. The survey is in alignment with the International Co-operative Programme on Assessment and Monitoring of Air Pollution Effects on Forests and thus uses the same survey protocol \citep{Eich2016}. The TCCI includes sampling points from the large-scale monitoring programme (level I) and longterm intensive monitoring areas (level II).  The background and description of these two parts of the survey is given in \cite{DaHeKoe2001,DeVries2003} and \cite{Eich2016}; see also Supplementary Material Section 1 for a detailed description of the survey protocol. 
The level I data are essentially yearly repeated measures on a regular spatial grid with different subsets of locations missing, depending on the year resulting in sampling points on either a $4\times4$, $8\times8$ or $16\times 16$ km grid. At each sampling grid point 24 sample trees with minimum height of 60 cm and a dominant and subdominant position within the forest stand
are randomly selected using a protocol ensuring good spatial coverage within a 50 m radius. The trees are permanently marked and re-assessed during subsequent surveys.
Trees that are removed are replaced by newly sampled trees. 
The level II sampling areas are of size 0.25 ha, predominantly in monoculture stands representing typical forest landscapes with the main species spruce, beech, fir, pine or oak. Only stands with trees of age 60 years or older were selected.  
We combine the level I (87\% of trees) and II (13\% of trees) of the years 1985 to 2013 data for our analysis and this results in 2254 sampling grid points. Figure~1 in the Supplementary Material shows that in 1985 and 1986 there was high proportion of grid points due to a   $4\times4$ km grid, compared to the following years.  Note that the median plant year is 1915, 1893, 1905, 1899 and 1904 for spruce, fir, pine, oak and beech respectively. Figure 2 in the Supplementary Material shows boxplots of the plant year by species. 
\begin{figure}
\vspace{-1cm}
\begin{center}
\includegraphics[scale=0.6,angle=0]{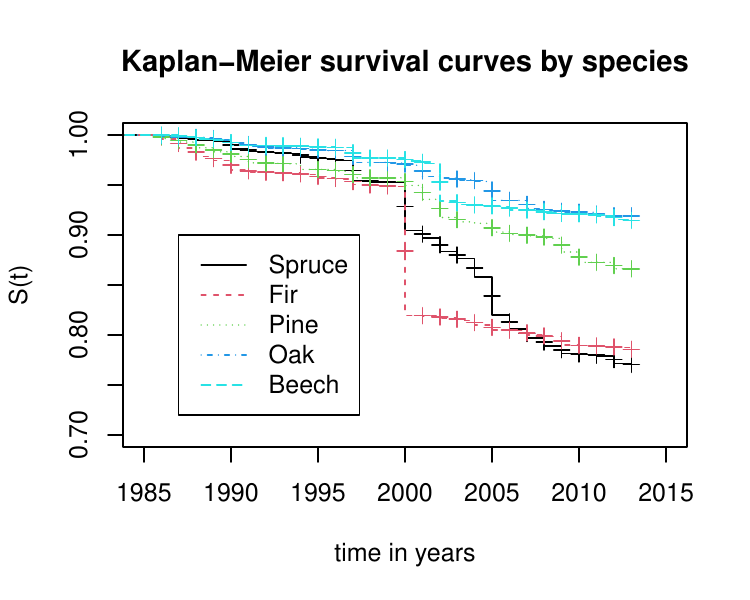}\includegraphics[scale=0.6,angle=0]{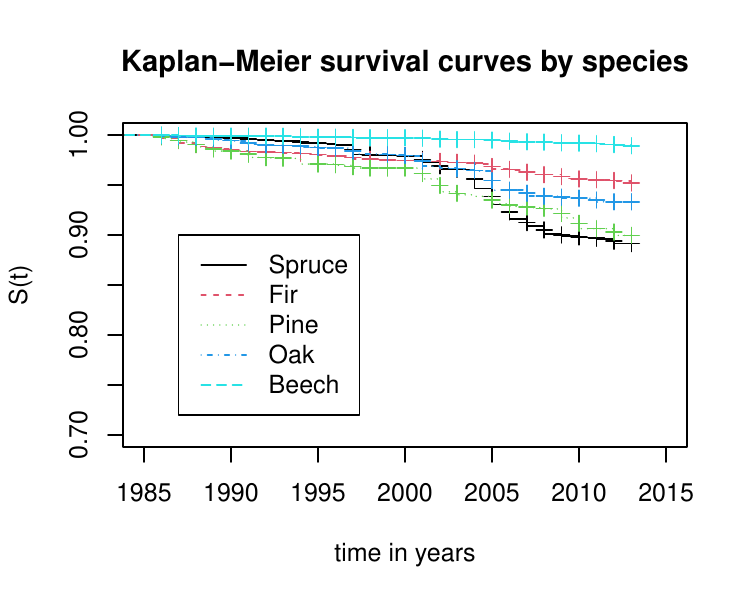}
\end{center}
\vspace{-1cm}
\caption{Kaplan-Meier survival curves by species. Curves are marked at each censoring time which is not also a death time. Scenario 1 left and Scenario 2 right.}
\label{km.graph}
\end{figure}
We have time-to-event data by tree and among the reasons why trees ceased to be observed are different type of events which are causes of death.
One type of event is death or damage due to storm. 
An argument for treating this as a relevant event is that not only climatic variables, describing windiness in terms of large area winter cyclones or small area convective (summer) storms, explain trends in storm damage. Additionally, the disposition of sites may be influenced by other characteristics sensitive to climate change effects such as water saturation in the soil due to potentially increasing winter precipitation or increasing winter temperatures with shorter periods of frozen soils. Both would indicate a tendency to increase storm damage due to reduced anchorage moments, even if the intensity and frequency of storms did not increase under climate change. 
The argument against including storm damage as a relevant event is that, although due to climate change there may be more storms, it is impossible to predict the storm path with the available environmental variables, and hence treating these as relevant events is not meaningful in our model set up. Also it is difficult to model this event simultaneously with mortality caused by dry-hot effects such as drought.
As neither argument is conclusive in the following we use two scenarios for defining relevant events:  Scenario 1 includes death due to storm damage and other reasons including lightning as relevant events, and  scenario 2 excludes storm damage.  
Events common to both scenarios include (see also Table~2 in the Supplemental Material): 
100\% defoliation,  
death due to unknown cause, 
harvest or death due to biotic damage and harvest or death due to abiotic damage which could not be attributed to storm damage (lightning, snow breakage etc.).
%
Trees are censored if the tree was
harvested during a regular, scheduled thinning, 
the tree has a broken crown, 
the tree was no longer observed as it did not fit in the criteria regarding social class, i.e. it changed social class. 
The tree is censored if the tree was no longer observed due to the site no longer being observed or the tree was dead due to storm damage (under scenario 2). 
Moreover in case the sampling point has not been visited for some number of years and at the next visit the tree has disappeared for unknown reason, the tree is censored. 
Also, if the tree is still alive at the end of the observation time in 2013 the tree is censored.

Table~\ref{tab:dead} gives the numbers of censored and dead trees for scenario 1 and 2.   In scenario 1 where storm damage counts as an event, fir has the highest percent of events (\%) followed by spruce and pine (\%). When storm does not count as an event under scenario 2, pine has the highest number of events (3\%).
\begin{table}[]
\caption{Number of censored and dead trees by species for scenario 1 and 2. Note that Table~2 in the Supplementary Material gives the detailed numbers and categories for event classification.}
\label{tab:dead}
\centering
\begin{tabular}{rrrrrr}
  \hline
 & Spruce & Fir & Pine & Oak & Beech \\ 
  \hline
  \multicolumn{6}{c}{Scenario 1}\\\hline
censored & 35687 & 6517 & 4612 & 3296 & 10321 \\ 
  dead & 1483 & 623 & 195 &  83 & 185 \\ 
 \% & 4.00 & 8.70 & 4.10 & 2.50 & 1.80 \\ 
   \hline
  \hline
  \multicolumn{6}{c}{Scenario 2}\\\hline
 & Spruce & Fir & Pine & Oak & Beech \\ 
  \hline
censored & 36604 & 6990 & 4661 & 3313 & 10475 \\ 
  dead & 566 & 150 & 146 &  66 &  31 \\
 \%  & 1.50 & 2.10 & 3.00 & 2.00 & 0.30 \\ 
 \hline
\end{tabular}
\end{table}
The Kaplan-Meier estimates of survival curves can serve as a descriptive reference for further model development and are shown in Figure~\ref{km.graph} for both scenarios 
of the main species. The curves are all very flat as the proportion of dead  trees is so small. Beech and oak have the flattest survival curves.  The year 1999/2000 stands out for fir because a lot of trees died due to the storm Lothar. 
Note that these Kaplan-Meier curves are adjusted for the fact that we have left truncation. The set of trees under study, or risk set, can grow or decrease with time. The risk set will decrease when a tree dies, but the risk set can grow since trees can be recruited into the study at any time. If a tree is recruited and we know its age, we know it was alive up until that point. We do not however, have knowledge of the trees in that area that died before the given tree was recruited.  
\subsubsection*{Explanatory variables}
We carried out a pre-selection of 60 original  explanatory variables  related to soil, climate and water budget characteristics from different sources based on pairwise correlation  and interpretability. If pairs of variables
have a strong association and are proxies of each other, we included the variable with fewer
missing values and/or the variable which is more useful in terms of identifying causes of tree
damage and predictor value ranges for critical conditions. The resulting variables  are listed in Table~\ref{tab:var}. Variables from the TCCI survey are year, northing, easting, altitude, planting year and defoliation. The variable defoliation is assessed by visual interpretation of current foliation relative to full foliation in the crown, estimated in $5\%$ classes for each individual tree using binoculars. 
Except for {\em year}, {\em defoliation} and {\em plantyear} which are available by individual tree,  these variables are available at sampling  site level only. 
The variables on soil properties are available once by sampling site location and are derived from the National Forest Soil Inventory carried out once between 2006 and 2008.  
The climate variables are available yearly, we use  different indicators for mean annual drought stress and aridity indices derived from daily weather variables (temperature, precipitation, solar radiation, humidity and wind speed) which were predicted on a 250m grid using statistical regionalization methods. The methodology for the generation of the weather grid data is described in \cite{DietWolKawo2017}. 
The yearly data on water budget was modelled using the process-oriented forest hydrological simulation model LWF-Brook90 \citep{HamKen2001}, which calculates the water budget of a one-dimensional, multi-layered soil profile with vegetation cover in daily resolution based on the model Brook90 \citep{FedVor2003}. LWF-Brook90 requires meteorological input data in daily resolution (precipitation, temperature, radiation, water vapour pressure, wind speed), derived from the 250m gridded weather  data described above.
\tiny
\begin{table}[]
\begin{center}
\caption{Variables used in the first stage of model selection using years 1985 to 2013. We state at which resolution the variables are available: year, location (the coordinates of sampling point) and tree. In bold the variable abbreviation is given. }
\label{tab:var}
\begin{tabular}{lll}
\hline
source& resolution&variable (unit)\\ \hline
TCCI& location, tree &{\bf defoliation}, {\bf plantyear} \\ 
& location & {\bf northing}, {\bf easting}, {\bf year}, {\bf altitude} (m)\\\hline
{\bf soil}&location &effective {\bf cation} exchange capacity  ($\mu$eq/q) \\
federal soil&&{\bf base sat}uration (\% x 10)\\ 
survey &&available {\bf water cap}acity (mm)\\ 
2006-2008&&soil {\bf depth} (m) \\ 
\hline
{\bf climate}&year, location&sum of {\bf glob}al{\bf  rad}iation (MJ/m$^2$) \\ 
weather model&& minimum temperature May ($^o$C) ({\bf tmin\_may}) \\ 
predictions&&average daily climatic {\bf water} {\bf balance} (mm)\\
1961- 2013&& {\bf aridity} index (cm/$^o$ C) after \cite{de1926areism}\\
&&{\bf stand}ardized {\bf prec}ipitation index for a  12 months period\\
&&sum of days with temp $>$ 20$^o$ C ({\bf tsum20})\\
&& julian day of {\bf budburst}\\\hline
{\bf water budget}&year, location &sum of {\bf trans}piration (mm)\\          
LWF-&&mean ratio between actual and potential transpiration {\bf transratio} \\
watercap90&&mean of relative plant available {\bf water storage} in root zone\\
&&{\bf min}inum of relative plant available \\
&&{\bf water storage} in root zone \\
&&number of days with relawat\_we\_avg $<$ 40 ({\bf waterstordays\_low40})\\
&&sum of water shortage for days with relawatwe $<$ 40 in (mm)\\
&&  ({\bf waterstorsum\_low40})\\
&&mean soil water potential in root zone  (hPa) ({\bf meansoilwat}) \\
&&number of days with soil water\\
&&  potential in root zone $<$ -1200 hPa ({\bf meansoilwatdays\_low1200}) \\
\hline
\end{tabular}
\end{center}
\end{table} 
\normalsize
\section{Modelling tree survival}
For investigating the relationship between mortality, environmental variables and defoliation we use a smooth additive Cox model, see e.g. \cite{hastie1987generalized}, which is an extension of the Cox model \citep{Cox1972,tian2005cox}. We model the survival time $t$ of tree $i$ with time $t$ equal to the calendar time at death. As we model survey data from 1985, this is time 0  in the usual sense of a Cox model.
This model relates  the hazard function, $h (t)$, to the baseline hazard function, $h_0 (t)$. A hazard function is a measure of risk or \textit{hazard} of some event occurring, in our case it is the risk of tree death. The baseline hazard function is the underlying hazard common to all individuals in the population. Some of our covariates are invariant in time (e.g. plantyear) and some others change in time, e.g. the water budget and climate variables, hence the following smooth additive Cox model for the hazard is used
\begin{equation} 
h(t) = h_0(t)\exp \left \{ {\sum}_k f_k(x_k) + {\sum}_j f_j(z_j(t))  \right \}  \exp \left \{ {\bf W} {\bf b} \right \} = h_0(t)\exp \left \{  \X(t)\bp   \right \}.
\label{mod.survtime}
\end{equation}
The funtions $f_k$ and $f_j$ are centred smooth functions of time invariant covariates, $x_k$ and time dependent covariates $z_j(t)$. The matrix ${\bf W}$ contains dummy variables relating to the random effects, which are i.i.d. Gaussian random effects with ${\bf b} \sim N({\bf 0},\sigma^2)$. The term  $\exp{\left ({\bf W} {\bf b}\right )}$ is often called a frailty term.  This includes a random effect for grid location taking care of dependence between neighbouring trees and implying the same level of correlation between any pair of trees at the same grid location. This covariance structure is often referred to as `compound symmetry'; see for example \cite{batespinheiro2007}.  To simplify notation we let $\X_i(t_i)$ be the combined model matrix with  row $\X_i(t_i)\ts$  for tree $i$ at event time $t_i$, covariates for random effects and basis functions evaluated at observations $i$ for covariates with smooth effects. The parameter vector $\bp$ contains all coefficients for $\X_i(t_i)$. 
Note that the baseline hazard changes by calendar time and hence can explain some temporal trends common to all trees driven by hidden explanatory variables. 
 We can adjust for tree age by including either age or plant year as a predictor.  

We model species separately for the following reasons. The species Norway spruce has shallow root systems and the other four species have not, with fir, pine and oak, having deep root systems and beech having heart shaped root systems. This means that spruce is more susceptible to drought, and the effect of drought can be modified by soil type, with drought having a worse effect on defoliation in sandy soil conditions. In addition, temporal correlation will be stronger in coniferous trees, as the trees hold needles from several previous years. The age distributions  of the different species also differ, with fir being the oldest followed by oak. Additionally, since storm risk, as one very important mortality risk for trees, is strongly correlated with tree height, due to mechanistic effects, survival models actually need to be species specific, since tree height growth varies strongly between species. This species-level differentiation would likely make parameter selection under grouping of tree species more unstable.
\subsection{Parameter estimation}
In the Cox model the baseline hazard, $h_0(t)$, is treated as a nuisance parameter, an arbitrary non negative function. This means that our data only includes information about which subject from the `at risk' group experienced the event at each time. Hence we only have information for estimating model parameters at times at which an event actually occurred. We also know the number of individuals at risk at each event time. The likelihood based on this informative data is the {\em partial likelihood}, $L(\bp)$ based on the linear predictor from \eqref{mod.survtime}.
Then
$$
L(\bp) = {\prod}_{i} \exp \{ \X_i(t_i)\bp\}/{\sum}_{l \in R(i)} \exp \{ \X_l(t_i)\bp\}
$$
where tree $i$ has event at $t_i$ and  $R(i)$ denotes the risk set. 
The parameter vector $\bp$ contains all coefficients for $\X_i(t_i)$. To control smoothness and dispersion of the random effect the penalized version of the log likelihood is optimised, so
\beq
\hat \bp = \underset{\bp}{\text{argmin}} -l(\bp) + {\frac{1}{2}} {\sum}_j \lambda_j \bp\ts  {\bf S}_j \bp \label{pl}
\eeq
with $l = \log L$ and the penalty parameters $\lambda_j$. The ${\bf S}_j$ is the penality matrix for smooth term or random effect $j$, and for a smooth term $\bp\ts  {\bf S}_j \bp$ measures 
 how wiggly it is. For example in the case of a cubic spline smooth  $\bp\ts  {\bf S}_j \bp$ is the  integrated squared second derivative of the smooth function it relates to. 
For the case of only having time invariant covariates it is possible to work directly with the partial likelihood given above. With careful iterative nesting of the risk set summations, the evaluation has $O(np^2)$ operations cost, with $p=\text{dim}(\bp)$.
In our case we also have time varying covariates and we use the equivalence of the partial likelihood to the likelihood of a Poisson model for pseudodata \citep{whitehead1980fitting}.
Each event time constitutes the interval of one year and at each interval a tree still at risk contributes its covariate and a binary indicator of whether its event occurs. By only including trees once they are first observed, we automatically deal with the left truncation of the  data; see also \cite{guo1993event}.   The linear predictor is as equation~\ref{mod.survtime} above but includes an extra factor with a level for each event time, required for the pseudodata likelihood to be equivalent to the partial likelihood.
Naively this approach has $O(n^2p^2)$ operations cost, but the approach of \cite{wood2017generalized} reduces the effective size (see below).  See also \cite{Wood17} and \cite{bender2018generalized} for details.

For estimating the smoothing parameters, the fact that the model can be seen as a Bayesian model is exploited. Then the  penalty is induced by a prior $\bp \sim N({\bf 0},{\bf S}_\lambda^-)$ with ${\bf S}_\lambda = \sum_j \lambda_j {\bf S}_j$. Note that ${\bf S}_\lambda$ contains components relating to the smooth effects and the random effects. The ${\bf S}_j$ for the random effect $\bf b$ is $\bf I$ and the respective $\lambda_j$ is $\sigma^{-2}$.   It follows that $\hat \bp$ in (\ref{pl}) is the maximum a posteriori estimate. ${\bf S}_\lambda^-$ is an appropriate pseudo inverse.
Fom this Bayesian perspective smooths can be seen as latent Gaussian random fields and hence in terms of estimation we don't need to differentiate between random effects and smooths (see e.g. \cite{kimeldorf1971some,Silv85}). Also if $\bm{{\cal I}} = \X^T{\bf W} \X$ is the information matrix,  with ${\bf W}=\mbox{diag}(w_i)$ and weights $w_i$ based on the Poisson density, for $n \to \infty$ and $p = o(n^{1/3})$ the posterior is  (see e.g. \cite{wood2016smoothing} for details) 
\beq
\bp |{\bf y},{\bm \lambda}\sim N(\hat \bp, (\bm{{\cal I}} + {\bf S}_\lambda)^{-1}) \label{dist}
\eeq
Using the above large sample approximation enables us in principle to estimate $\bm \lambda$ to maximize the marginal (partial) likelihood
$$
f({\bf y}|{\bm \lambda}) = f({\bf y},\hat \bp|{\bm \lambda})/f(\hat \bp|{\bf y},{\bm \lambda})
= L(\hat \bp)f(\hat \bp|{\bm \lambda})/f(\hat \bp|{\bf y},{\bm \lambda})
$$
Using (\ref{dist}) to approximate $f(\hat \bp|{\bf y},{\bm \lambda})$ yields the  Laplace approximated maximum likelihood (LAML). Here the vector $\bf y$ refers to the binary indicator event indicator explained above.

In practice the LAML  is optimised  w.r.t. ${\bm \rho} = \log {\bm \lambda} $ by updating $\bm \rho$ to increase the marginal likelihood of the working linear mixed model at each step of a penalised penalized iteratively re-weighted least squares iteration scheme for estimating $\bp$.  
This method can be parallelised, and made very efficient by marginal discretization of covariates (or exploiting natural discretization) \citep{wood2017generalized}, thereby mitigating the $O(n^2p^2)$ cost. Due to the large number of parameters and large sample sizes we use this efficient estimation implemented in the function $\tt bam$ in the R package {\tt mgcv} in R.
\section{Model selection}
\subsection{Standard tools}
\label{standard}
Considering standard GAM model selection tools we can obtain approximate p-values for testing smooths or random effects for equality to zero \citep{wood2013simple,wood2012p}. A generalised $\mbox{AIC} = -2l(\hat \bp) + 2 \tau$ can be computed using $\tau$, the { effective degrees of freedom} for a penalized model 
$$
\tau = \text{trace}\{{\bm V}_\beta \bm{{\cal I}} \}.
$$
where ${\bf V}_\beta = (\bm{{\cal I}} + {\bf S}_\lambda)^{-1}$ is the approximate Bayesian posterior correlation matrix. 
\cite{greven2010behaviour} show that this can sometimes perform poorly (selecting larger more complex models) due to neglect of smoothing parameter uncertainty. 
But we can correct for smoothing parameter uncertainty using the techniques of \cite{wood2016smoothing}.
The simplest correction adds a correction  to the Bayes correlation matrix  ${\bf V}_\beta^\prime = {\bm V}_\beta +{\bf V}^\prime$, where  
${\bf V}^\prime={\bf J}  V_{\bm \rho} {\bf J}^T$ and
${\bm V}_\beta = (\bm{{\cal I}} + {\bf S}_\lambda)^{-1}$, 
with ${\bf J}_{ij}= \frac {\delta\hat \beta_i}{\delta {\bm \rho}_j}$ 
and ${\bf V}_\rho$ can be estimated by the inverse Hessian of the approximated maximum likelihood optimised to find  ${\bm \rho} = \log {\bm \lambda}$. See \cite{wood2016smoothing}, who also provide a better approximation. 

So, using either AIC or p-values it is possible to carry out backward selection, but this becomes computationally expensive for our model with a large number of potential predictors.

It is also possible to carry out integrated backward selection by adding extra selection penalties (double penalty approach \citep{wood2011fast, marra2011practical}). 
This method adds a penalty to the coefficients of each smooth, penalizing the functions in the null space of its smoothing penalty, so that they can in principle be penalized out of the model. Although the estimation of $\bm \lambda $ already performs a lot of model selection,  $\lambda_j \to \infty$ does not usually remove a term altogether. This is because, for example, with penalty $\int f_j^{\prime\prime}(x)^2$, the $\hat f_j(x) \to \alpha_0 + \alpha_1 x$ (the null space) as $\lambda_j \to \infty$.
Thus each $f_j$ is given an extra penalty, penalizing the null space of its smoothing penalty. So $\bm \lambda$ estimation now does integrated model term selection. However large numbers of predictors are again challenging, or even computationally infeasible.  
\subsection{Gradient boosting with integrated selection}
\label{sec:boogam}
To avoid the excessive cost of stepwise model selection methods, requiring many large model sets, or of penalised selection 
requiring very many model fits with excessive numbers of penalties, we propose an efficient novel stepwise method. 
The proposal is to combine efficient forward selection via boosting, with relatively economical backward selection via selection penalties. 

Component-wise gradient boosting \citep{HofMayNik2014, HotBueKne2010,BueHot2007} offers very efficient forward model selection, because it only requires uni-variate models fitted to the gradient of the likelihood. 
We first illustrate the idea of smoothing by boosting in one dimension, with least squares loss: 
\begin{enumerate}
\item Construct a low degree of freedom linear `base smoother' $\hat {\bf y} = {\bf Ay}$, where ${\bf A} = \X(\X\ts \X + \lambda_\text{big} {\bf S})^{-1} \X\ts$ and $\X$ is the model matrix for a univariate smooth containing the basis functions evaluated at the covariate values. $\lambda_\text{big}$ is the penalty or smoothing parameter which is fixed and large.
\item Initialize $\hat {\bf f} = {\bf 0}$ and then iterate $\hat {\bf f} \leftarrow \hat {\bf f} + {\bf A}({\bf y} - \hat {\bf f})$. This is done until the final stopping iteration. The optimal final stopping iteration can be estimated using cross-validation or AIC. 
\end{enumerate}
This idea is used in gradient boosting with integrated selection
where we consider a model with a log likelihood $l$ and multiple smooth terms, $f_j$, in a linear predictor $\bm \eta$.
We set up base smoothers (hat matrix ${\bf A}_j$) for each $f_j$ potentially in the model and then iterate the following steps \citep{schmid2008boosting,mayr2012generalized}. Note that we present two options for the step length (a) $\alpha$ to be estimated and (b) $\alpha$ is fixed as recommended in \cite{BueHot2007}.
\begin{enumerate}
\item Compute gradient $e_i = -\ildif{l}{\eta_i}$ at observation $i$.
\item For all $j$ compute $\tilde {\bf f}_j = {\bf A}_j {\bf e}$ and find $\hat \alpha_j = \text{argmax}_\alpha l({\bm \eta} + \alpha \tilde {\bf f}_j)$. 
\item \begin{enumerate}
       \item [(a)]  Find $k = \text{argmax}_j l({\bm \eta} + \hat \alpha_j \tilde {\bf f}_j)$ or

       \item [(b)] Find  $k = \text{argmax}_j l({\bm \eta} +  \alpha \tilde {\bf f}_j)$ with $\alpha= constant$.
\end{enumerate}
\item Set ${\bm \eta} \leftarrow {\bm \eta} + \hat \alpha_k \tilde {\bf f}_k$, and add term $k$ to the set of selected terms. 
\end{enumerate}
This is a very efficient forward selection method particularly suited to the scenario with a large number of predictors, but suffers from being `forward only', the need to specify a stopping criterion and the lack of a full inferential framework for the resulting model. Penalized regression offers an efficient means for
backward selection, but is computationally costly when using GAMs with a large number  of candidate predictors. Our approach is therefore
to combine the two in a `boost forward penalise backward' scheme using the selection penalties as described in section~\ref{standard}:

We start with a minimal model of terms that {\em must} be included (based on arguments described in section~\ref{modsel:strat}), and create base smoothers for all other terms that are {\em potentially} in the model. Then iterate the following steps
\begin{enumerate}
\item Given the current model, run a few steps of gradient boosting to identify terms to include in the model from those not currently in.
\item Add the identified terms to the model and refit with selection penalties on all terms not originally in the base model. 
\item Drop any terms that are penalized out of the model from the current model. 
\end{enumerate}

\hspace{0.5cm} Repeat the above steps until the fit stops changing.

This approach is very simple, and has several advantages: Forward and backward steps are much more efficient than forward or backward selection by direct penalized MLE, as we eliminate the need to re-fit a model including each candidate predictor at each forward step. The stopping rule, inference and no-way-back problems of boosting are avoided.  We investigate the selection performance of the proposed gradient boosting with integrated selection in comparison to integrated backward selection  by a simulation study in section~2 of the Supplementary Material. Based on the best performance of false positive rates in the simulation study (section 2 Supplementary Material) we use method (b) gradient boosting with integrated selection with fixed step length for model selection here. In this application forest planning relies on the variable selection results and hence we need to avoid false positives.
\subsection{Model selection strategy}
\label{modsel:strat}
Plant year ({\em plantyear}) is always included as a predictor in the model selection due to systematic differences between level I and II and in order to avoid bias in estimates \citep{Pencina2007}. Also we always include a random $w_i$ effect due to location to take account of spatial correlation between trees within a survey grid point. The simplest model we consider is the following
\begin{equation} 
\mbox{min\_model:}~~\log\frac {h_{i}(t)}{h_0(t)} = f_1({\tt plant year}_{i}) + w_i\ts {\bf b} 
\label{mod.min}
\end{equation}
for tree $i$ and $w_i$ is an indicator vector  for grid location of tree $i$ and ${\bf b}$ as Gaussian random effect for location as described in equation~\ref{mod.survtime}.
We carry out the `gradient boosting with integrated selection' described in section~\ref{sec:boogam} with the minimal model~\ref{mod.min} selecting from variables shown in Table~\ref{tab:var}. Note that for most of the time varying variables we also used their lagged values from the previous year. 
We carry out this selection for all of the five species each for scenario 1 and 2 and the general form of the selected models is: 
\begin{equation} 
\log\frac {h_{i}(t)}{h_0(t)} = \mbox{min\_model} + \sum_k f_k(x_{ik}) + \sum_l f_l(z_{il}(t)) 
\label{mod:full}
\end{equation}
The term  $\sum_k f_k(x_{ik})$ contains functions of all the selected time invariant variables (soil and altitude) and the term $\sum_l f_l(z_{il}(t))$ contains all the selected time variant variables (climate and waterbudget). We fit separate models for the five tree species.

In order to see whether defoliation is a proxy for the environmental variables we further fit the following models:
\begin{equation} 
\log \frac {h_{i}(t)} { h_0(t)} =  \mbox{min\_model} +  f_3({\tt defol}_i)
\label{mod:defol}
\end{equation}
and
\begin{equation} 
\log \frac {h_{i}(t)} { h_0(t)} =  \mbox{min\_model} +  f_3({\tt defol}_i) + \sum_k f_k(x_{ik}) + \sum_l f_l(z_{il}(t)).
\label{mod:fulldefol}
\end{equation}
where the term  $\sum_k f_k(x_{ik})$ contains functions of all the selected time invariant variables (soil and altitude) and the term $\sum_l f_l(z_{il}(t))$ contains all the selected time variant variables (climate and waterbudget) from model~\ref{mod:full}.

\subsection*{Defoliation threshold estimation}
\begin{figure}[!h]
\begin{center}
\includegraphics[angle=0,scale=0.25]{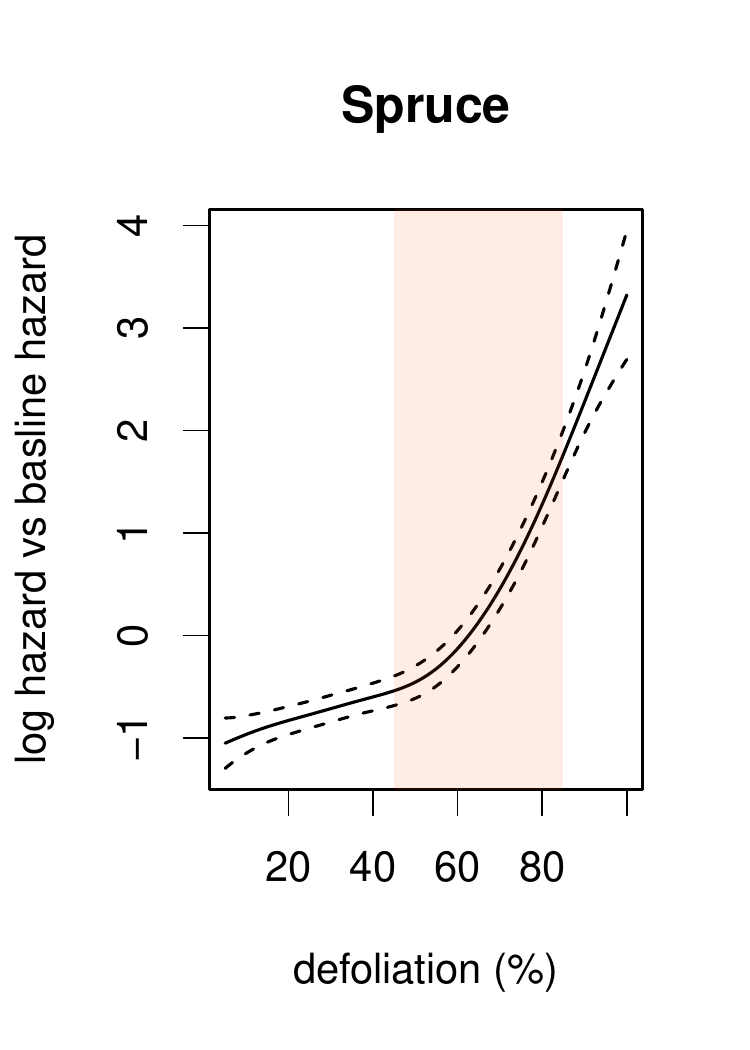}\includegraphics[angle=0,scale=0.25]{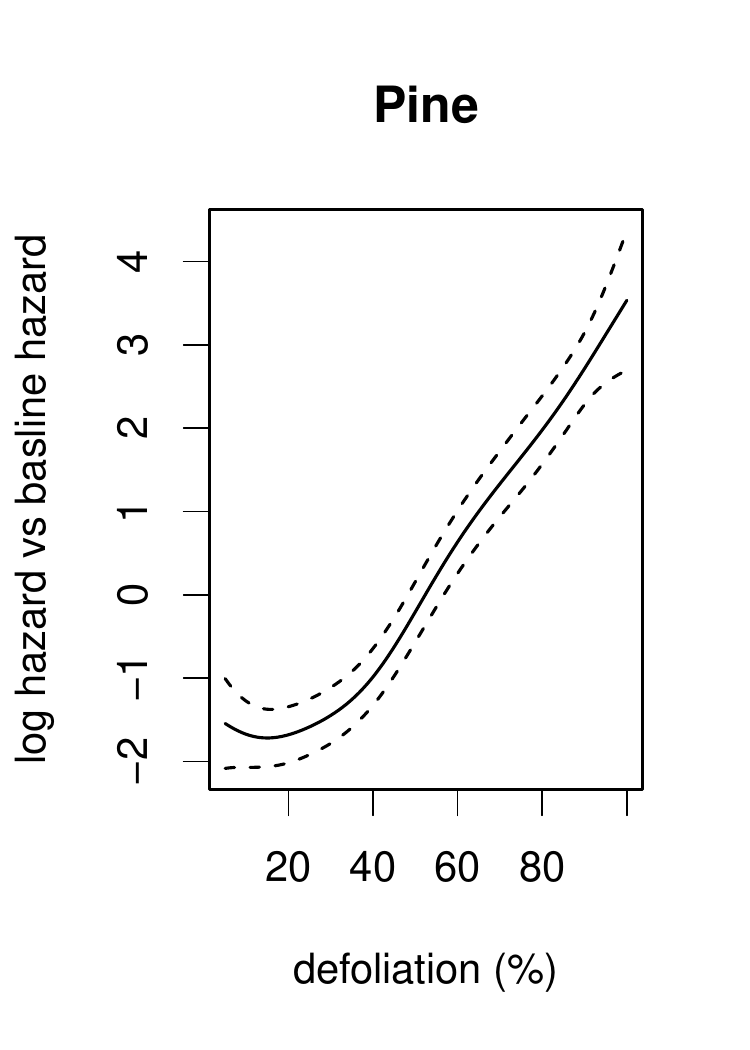}\includegraphics[angle=0,scale=0.25]{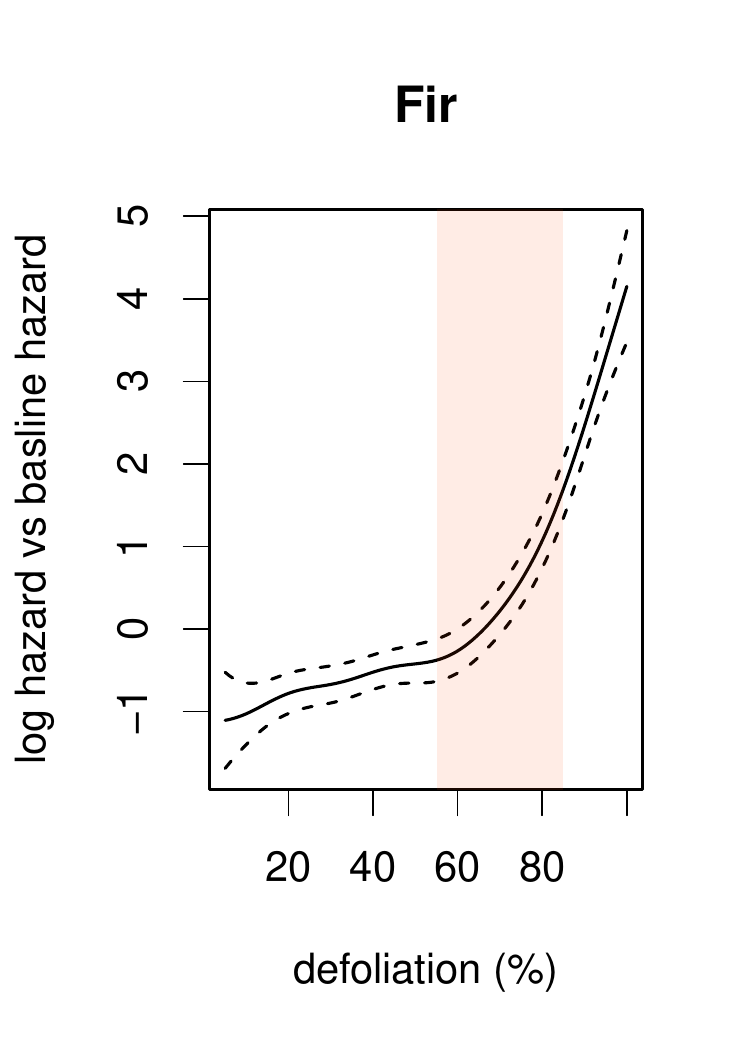}\includegraphics[angle=0,scale=0.25]{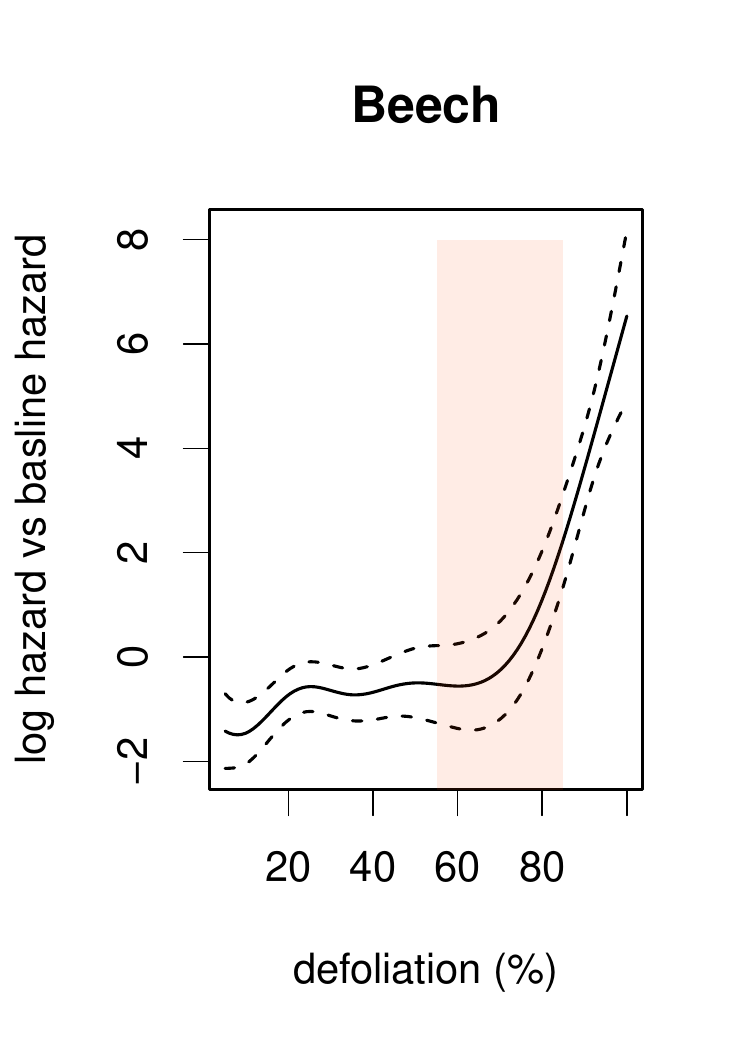}\includegraphics[angle=0,scale=0.25]{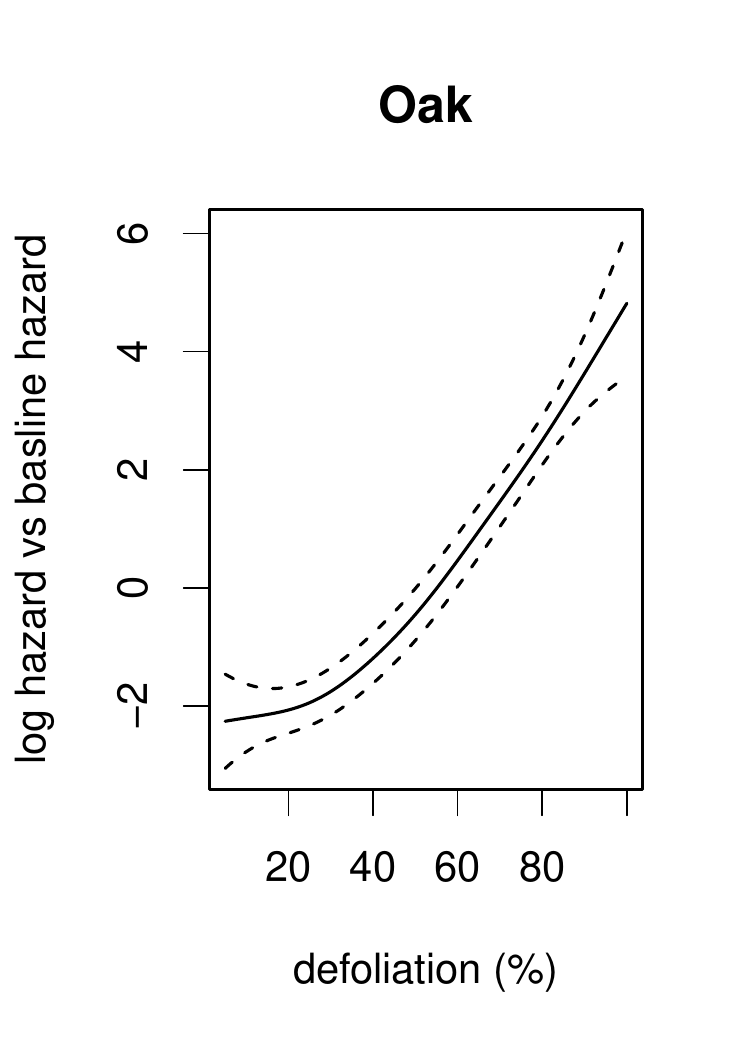}\\
\includegraphics[angle=0,scale=0.25]{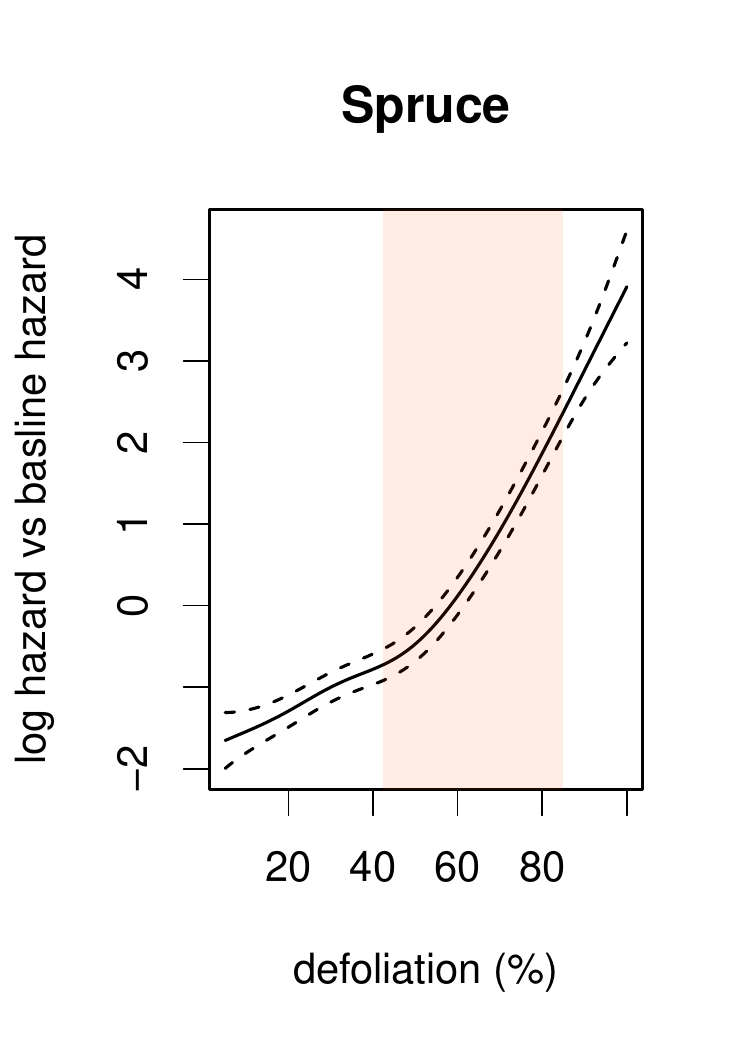}\includegraphics[angle=0,scale=0.25]{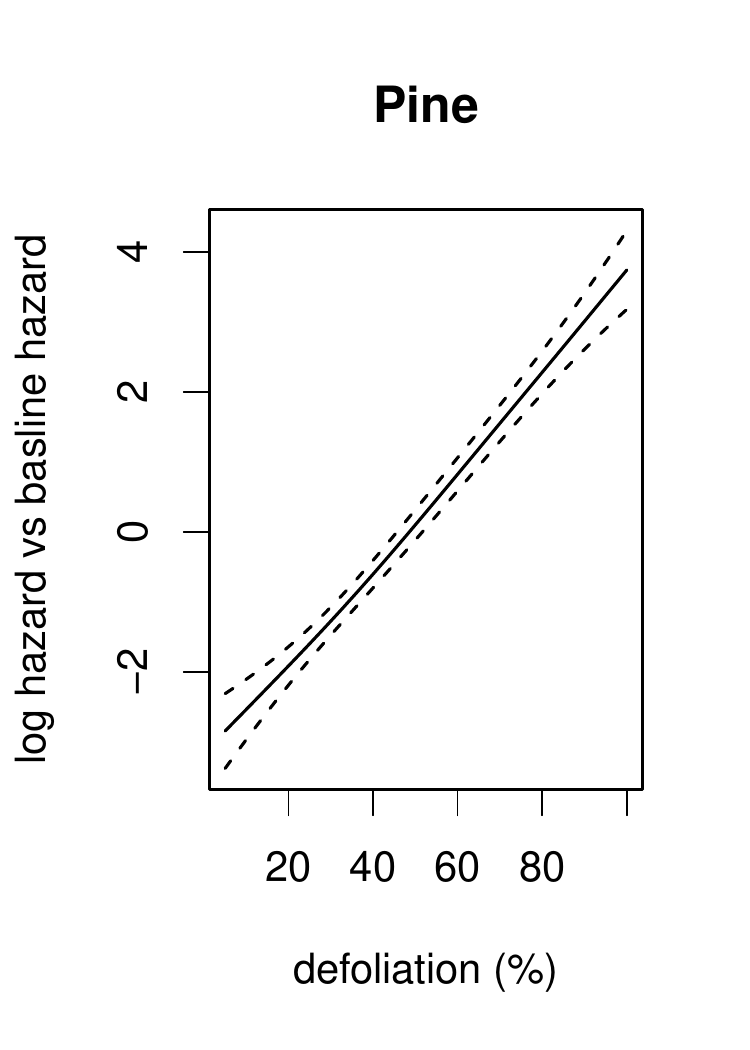}\includegraphics[angle=0,scale=0.25]{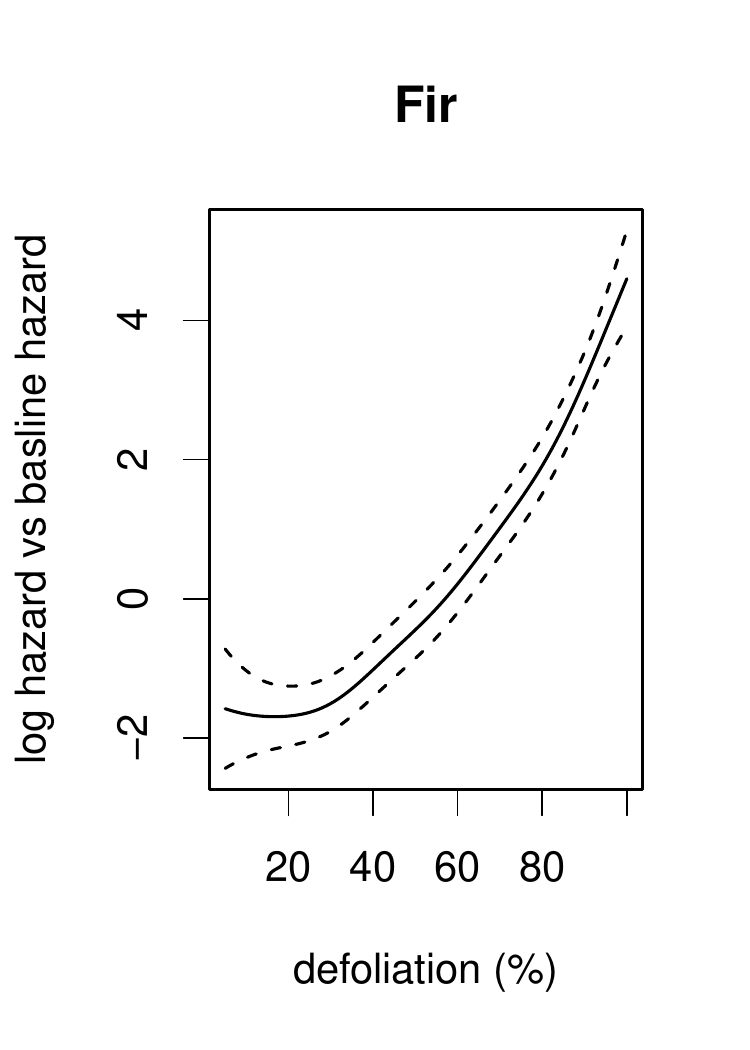}\includegraphics[angle=0,scale=0.25]{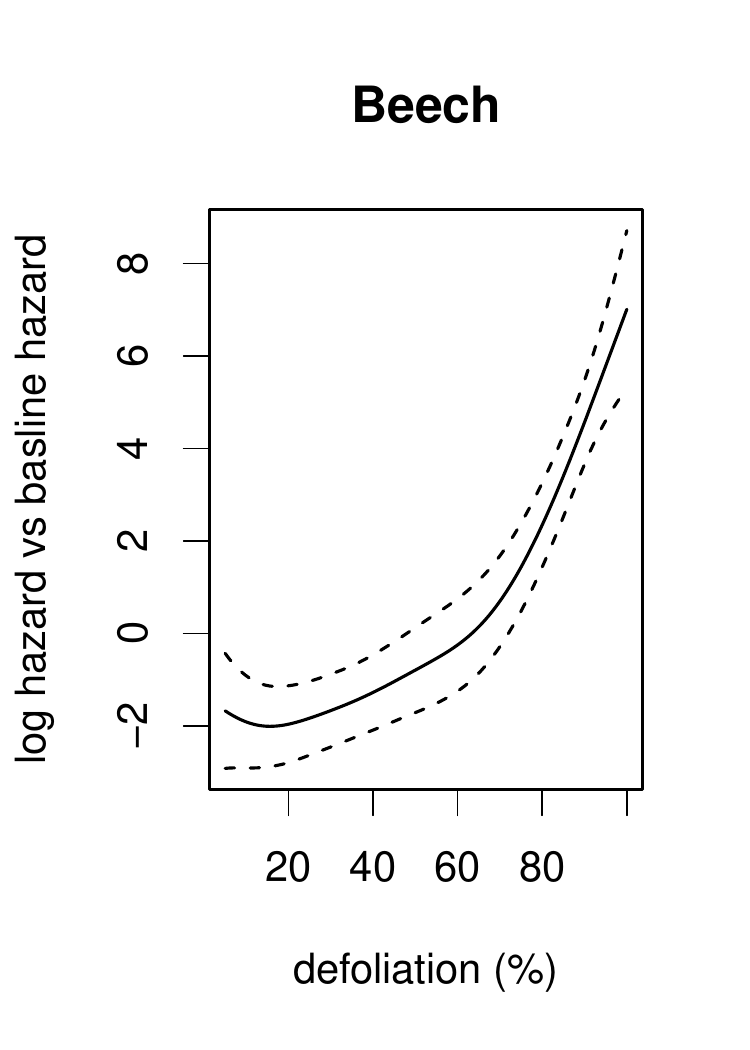}\includegraphics[angle=0,scale=0.25]{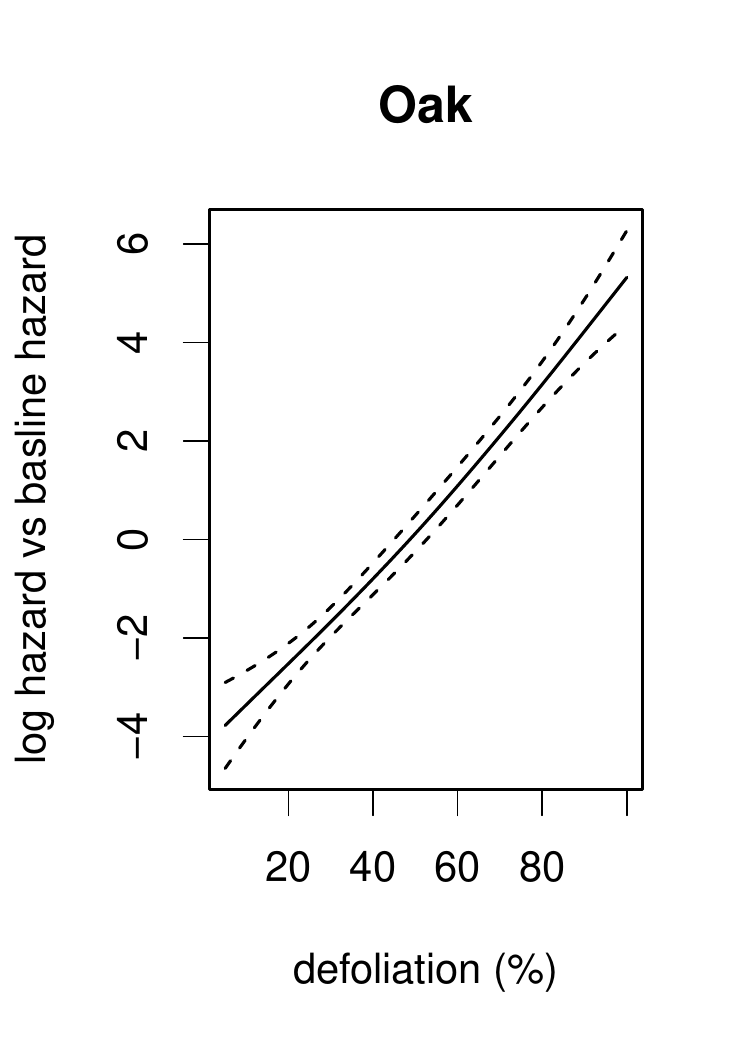}
\end{center}
\vspace{-0.5cm}
\caption{Effect of defoliation in previous year (from full model) Scenario 1 (top) and 2 (bottom). The shaded areas indicate the Baysian credible intervals of the defoliation thresholds ($\psi$). For graphs without a shaded area a clear threshold could not be found. }
\label{fig:loghazard}
\end{figure}

Turning to the key question of  defoliation. The effect of defoliation on the log hazard appears to be non-linear as seen in Figure~\ref{fig:loghazard} which shows the estimated smooth function of the log hazard of defoliation using the selected model with defoliation added (\ref{mod:fulldefol}). 
It appears that spruce trees with defoliation of around 60$\%$ or less have a different risk to those with a higher percentage defoliation. Identifying at what percentage defoliation the hazard increases more steeply will give a threshold value for when a tree would be most at risk of death. 

To estimate the change in slope seen in Figure~\ref{fig:loghazard} we estimate the second derivative  of $f_3(\tt{defol})$ along a grid of {\tt defol} values with finite difference interval $h$: 
$$
\hat f_3^{\prime\prime}({\tt defol}) = \frac {\hat f_3({\tt defol}+h) - 2 \hat f_3({\tt defol}) + \hat f_3({\tt defol}-h)}{h^2} .
$$ 
and with $\hat f_3(\tt{defol}) = \bf X^{\tt{defol}} \hat \bp^{\tt{defol}}$,
where $\hat \bp^{\tt{defol}}$ are the estimated 
coefficients relating to the basis functions of defoliation $\bf X^{\tt{defol}}$. Then 
the estimate of the second derivative is:
$$ 
\hat f_3^{''}({\tt defol}) = h^{-2} \left ( \bf X^{\tt{defol}+h} - 2 X^{\tt{defol}} +  X^{\tt{defol}-h} \right )  \hat \bp^{\tt{defol}} = \bf D \hat \bp^{\tt{defol}}
$$
and then the estimated breakpoint is
$
\hat \psi = \underset{\tt defol}{\text{argmax}} ~~\hat f_3^{''}(\tt{defol}).
$
For  Bayesian credible intervals of $\psi$ we draw samples $\tilde \bp^{\tt{defol}}$  from the posterior of $\bp$ (equation~\ref{dist}). For each of the samples we obtain $\tilde{f_3}^{''}(\tt{defol})$ and $\tilde \psi$.  The lower and upper 95\% quantiles constitute the Bayesian confidence intervals of $\psi$. If these intervals cover the entire range of the variable defoliation (0 - 100\%) then there is no evidence of a threshold. 
\normalsize
\begin{figure}[!h]
\vspace{-1cm}
\begin{center}
\includegraphics[angle=0,scale=0.9]{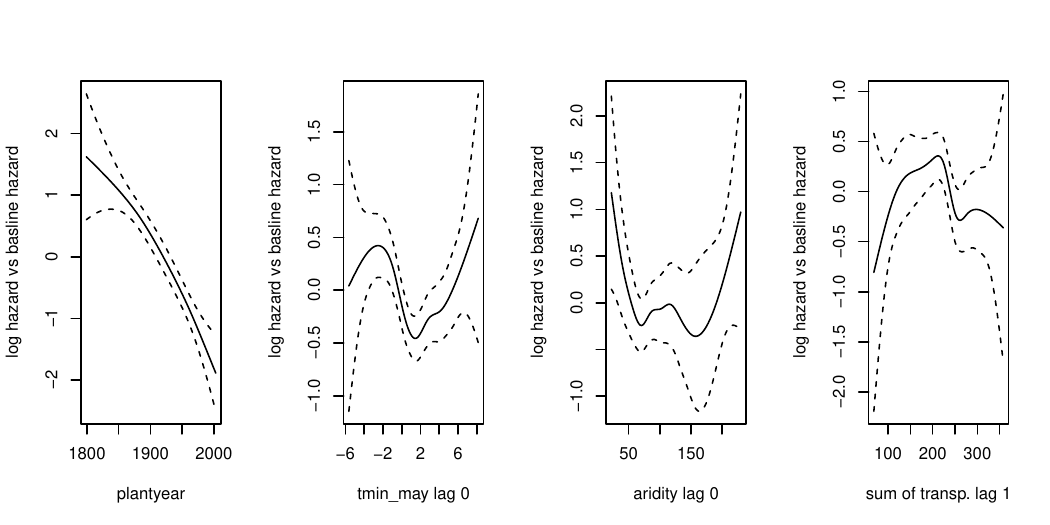}
\caption{Selected smooth effects for spruce for Scenario 1 with 95\% confidence intervals.}
\label{fig:smoothsel}
\end{center}
\end{figure}
\section{Results and interpretation}
Table~\ref{Tab:selvar} shows the selected variables of the 5 models fitted to spruce, fir, pine, beech and oak data for scenario 1 and 2.  
Note that the variable plant year and the random effect are in the minimal model~\ref{mod.min} and hence always contained in models.
The variable `mininum temperature May' (tmin$\_$may$\_$l0) was selected 7 times, in all the species except for oak. 
The table also shows that for  oak the climate variables were not as often selected as for the other species.   Many of the water budget variables are selected for the spruce model as well and, to a much lesser extent, in the other species models. Due to the fact that many of the explanatory variables are correlated because they measure similar attributes, it is problematic to interpret the estimated smooth effect curves. We have therefore
 picked out some explanatory variables with plausible effects for the spruce model under scenario 1 shown in (Figure~\ref{fig:smoothsel}). The effect of plant year is negative for all species escept for oak where it is not significant in scenario 1. In scenario 2 only the model for spruce has a signifcant negative effect of plant year.   We show selected smooths for both scenarios and all species in Figures~4 to 12 in the Supplementary Material. 
\begin{singlespace}
\tiny
\begin{table}[]
\centering
\caption{Model selection results. The table gives the variables selected for the survival models of the five main species under scenario 1 (including storm damage as event) and 2 (not including storm damage as event). For time varying variables the extension l0 and l1 denote current year or lag one year. Note that the variable plant year and the random effect are forced into all models.}
\label{Tab:selvar}
\centering
\begin{tabular}{rlr|ll|ll|ll|ll|ll|}
  \hline
  & & & \multicolumn{2}{c}{Spruce} & \multicolumn{2}{c}{Fir}  &\multicolumn{2}{c}{Pine} & \multicolumn{2}{c}{Beech}&\multicolumn{2}{c}{Oak}\\ \hline
  & variable & no sel. &  1 &  2 &  1 &  2 &  1 &  2 &  1 &  2 &  1 &  2 \\ 
  \hline
1 & plantyear & 10 & x & x & x & x & x & x & x & x & x & x \\ 
  2 & altitude & 1 &  &  &  &  &  &  & x &  &  &  \\ 
  3 & depth & 1 &  &  &  &  &  &  &  &  & x &  \\ 
  4 & cation & 1 &  &  &  &  &  &  & x &  &  &  \\ 
  5 & basesat & 0 &  &  &  &  &  &  &  &  &  &  \\ 
  6 & watercap & 2 &  &  &  &  &  & x &  & x &  &  \\ \hline
  7 & globrad\_l1 & 1 & x &  &  &  &  &  &  &  &  &  \\ 
  8 & tmin\_may\_l0 & 7 & x & x & x & x & x & x & x &  &  &  \\ 
  9 & water\_balance\_yl0 & 3 &  &  &  &  &  &  & x &  & x & x \\ 
  10 & water\_balance\_yl1 & 2 & x &  & x &  &  &  &  &  &  &  \\ 
  11 & aridity\_l0 & 5 & x & x & x &  &  & x & x &  &  &  \\ 
  12 & arididty\_l1 & 0 &  &  &  &  &  &  &  &  &  &  \\ 
  13 & standprec\_l0 & 1 & x &  &  &  &  &  &  &  &  &  \\ 
  14 & standprec\_l1 & 4 & x & x &  &  & x & x &  &  &  &  \\ 
  15 & tsum20\_l0 & 2 &  &  &  &  &  &  & x & x &  &  \\ 
  16 & tsum20\_l1 & 2 & x &  &  &  & x &  &  &  &  &  \\ 
  17 & budburst\_l0 & 2 &  &  &  & x &  & x &  &  &  &  \\ 
  18 & budburst\_l1 & 3 & x &  & x &  &  &  & x &  &  &  \\ \hline
  19 & tran\_l0 & 2 & x &  &  &  &  &  & x &  &  &  \\ 
  20 & tran\_l1 & 4 & x &  & x &  &  &  &  &  & x & x \\ 
  21 & tranratio\_avg\_l0 & 2 & x &  &  &  &  & x &  &  &  &  \\ 
  22 & tranratio\_avg\_l1 & 1 &  & x &  &  &  &  &  &  &  &  \\ 
  23 & water\_storage\_l0 & 2 & x &  &  &  &  &  & x &  &  &  \\ 
  24 & water\_storage\_l1 & 1 &  &  & x &  &  &  &  &  &  &  \\ 
  25 & minwater\_storage\_l0 & 1 & x &  &  &  &  &  &  &  &  &  \\  
  26 & minwater\_storage\_l1 & 3 & x & x &  &  & x &  &  &  &  &  \\ 
  27 & waterstorsum\_lower40\_l0 & 1 & x &  &  &  &  &  &  &  &  &  \\ 
  28 & waterstorsum\_lower40\_l1 & 2 & x &  &  & x &  &  &  &  &  &  \\ 
  29 & defsum\_relawatwe\_lower40\_l0 & 0 &  &  &  &  &  &  &  &  &  &  \\ 
  30 & defsum\_relawatwe\_lower40\_l1 & 2 &  &  & x &  &  &  & x &  &  &  \\ 
  31 & meansoilwat\_l0 & 2 &  &  & x &  &  &  & x &  &  &  \\ 
  32 & meansoilwat\_l1 & 2 & x & x &  &  &  &  &  &  &  &  \\ 
  33 & meansoilwatdays\_low1200\_l0 & 1 &  &  &  &  &  &  & x &  &  &  \\ 
  34 & meansoilwatdays\_low1200\_l1 & 4 &  &  &  &  & x &  &  & x & x & x \\ 
   \hline
\end{tabular}
\end{table}
\end{singlespace}
\normalsize
The results in Table~\ref{tab:modsel} can be used to investigate whether defoliation is a proxy for the environmental variables. As the sample sizes, here effectively the number of events, are relatively small for some of the species given the model parameters we use the AIC for this comparison.  
Comparing model~\ref{mod:full} with model~\ref{mod:defol} in Table~\ref{tab:modsel} shows that that under scenario 1 for the species spruce and beech the AIC for model~\ref{mod:full} is lower than for model~\ref{mod:defol}. This implies that  for spruce and beech the environmental variables (on soil, climate and water budget) explain more than defoliation of the previous year. Comparing model~\ref{mod:full} with model~\ref{mod:fulldefol} shows that the environmental variables explain more than defoliation of the previous year, but don't entirely replace defoliation.  
 For the species pine, fir and oak it was the other way round, defoliation of the previous year explained more than the environmental predictors, implying that it is unlikely that defoliation is a proxy for environmental predictors. For scenario 2 which does not count death due to storm as a relevant event (Table~2 in Supplementary Material) model~\ref{mod:defol} is always better than model~\ref{mod:full} for all species. But for all species and under both scenarios models with the defoliation effect added to selected environmental predictors (model~\ref{mod:fulldefol})  
 always have the lowest AIC.  
\begin{singlespace}
\begin{table}[]
\centering
\caption{Model results for scenario 1. Model selection as described in Section~\ref{modsel:strat}. The abbreviation 'min\_mod' is the minimal model given in equation  (\ref{mod.min}) with a smooth function of plant year and random effect due to location. 'envpred' stands for selected environmental predictors and 'defol' for defoliation of the previous year.  In bold is the best model for AIC and deviance explained. Note that the number of events is smaller than the reported in Table~\ref{tab:dead} due to missing values in the environmental variables. The sample sizes of the pseudo-data are 199976, 61173, 57302, 29188 and 17242 for spruce, beech, fir, pine and oak respectively.}
\label{tab:modsel}
\begin{tabular}{rrrrrr}
  \hline
 species and model & events & edf & dev. expl. & AIC & REML \\ 
  \hline
spruce minmod  (\ref{mod.min}) - frailty & 1387 & 34.30 & 0.19 & 14037.04 & 189394.98 \\ 
spruce  minmod (\ref{mod.min}) & 1387 & 375.47 & 0.42 & 11494.04 & 188258.69 \\ 
spruce  minmod+envpred (\ref{mod:full}) & 1387 & 460.11 & 0.47 & 11023.80 & 188053.85 \\ 
spruce  minmod+defol  (\ref{mod:defol})& 1387 & 376.57 & 0.44 & 11287.20 & 188155.89 \\ 
spruce  minmod+defol+envpred  (\ref{mod:fulldefol})& 1387 & 456.04 & {\bf0.48} & {\bf10841.33} & 187958.08 \\ 
    \hline
  \hline
beech minmod  (\ref{mod.min}) - frailty & 175 & 36.91 & 0.21 & 2036.93 & 57037.45 \\ 
beech  minmod (\ref{mod.min}) & 175 & 125.00 & 0.42 & 1798.68 & 56926.49 \\ 
beech  minmod+envpred (\ref{mod:full}) & 175 & 147.22 & 0.50 & 1681.59 & 56846.02 \\ 
beech  minmod+defol   (\ref{mod:defol})& 175 & 128.58 & 0.46 & 1721.44 & 56889.53 \\ 
beech  minmod+defol+envpred  (\ref{mod:fulldefol})& 175 & 153.40 &{\bf 0.55} & {\bf1595.31} & 56806.96 \\ 
   \hline
  \hline
fir minmod (\ref{mod.min}) - frailty &529  & 36.13 & 0.34 & 4384.79 & 54302.46 \\ 
fir   minmod (\ref{mod.min}) & 529 & 153.43 & 0.57 & 3502.41 & 53894.58 \\ 
fir   minmod+envpred (\ref{mod:full})& 529 & 174.41 & 0.61 & 3351.26 & 53810.59 \\ 
fir   minmod+defol  (\ref{mod:defol}) & 529 & 149.98 & 0.60 & 3332.06 & 53809.36 \\ 
fir   minmod+defol+envpred  (\ref{mod:fulldefol}) & 529 & 168.60 & {\bf 0.64} & {\bf 3176.36} & 53722.66 \\ 
   \hline
  \hline
 pine minmod  (\ref{mod.min}) - frailty & 144 & 27.88 & 0.09 & 1744.04 & 27525.77 \\ 
 pine   minmod (\ref{mod.min})& 144 & 113.90 & 0.29 & 1600.17 & 27473.00 \\ 
 pine   minmod+envpred (\ref{mod:full}) & 144 & 127.53 & 0.33 & 1572.44 & 27450.07 \\ 
 pine   minmod+defol  (\ref{mod:defol})& 144 & 97.96 & 0.39 & 1413.78 & 27373.89 \\ 
 pine   minmod+defol+envpred  (\ref{mod:fulldefol})& 144 & 110.37 & {\bf 0.43} & {\bf 1391.47} & 27354.43 \\ 
   \hline
  \hline
oak minmod  (\ref{mod.min}) - frailty & 67 & 20.00 & 0.08 & 861.38 & 16180.78 \\ 
 oak  minmod (\ref{mod.min})& 67 & 59.96 & 0.23 & 830.78 & 16166.82 \\ 
oak   minmod+envpred (\ref{mod:full})& 67 & 59.16 & 0.25 & 819.83 & 16150.73 \\ 
 oak  minmod+defol  (\ref{mod:defol})& 67& 43.80 & 0.39 & 681.96 & 16088.24 \\ 
oak   minmod+defol+envpred  (\ref{mod:fulldefol})& 67 & 44.04 & {\bf 0.40} & {\bf 679.72} & 16077.29 \\ 
   \hline
\end{tabular}
\end{table}
\end{singlespace}
Figure~\ref{fig:loghazard} shows the estimated hazard rate on the logarithm scale as a function of defoliation of the previous year for scenario 1 based on the full model. We are interested to see whether there is a value of defoliation at which the tree is increasingly more likely to die. It is apparent that with increasing defoliation, the tree's hazard of death increases. Also, for spruce, fir and beech there is an indication of a threshold above which the hazard increases sharply. 

We show the 95\% Bayesian confidence interval of the threshold value shaded in orange. 
In the case of spruce, this confidence interval is between 45 and 85\%.  In the case of fir it is between 55 and 85\% and for beech it is between 55 and 85\%. 
It is interesting to see that for beech we have two steps of the hazard rate, initially with low defoliation up to about 20\% the log hazard rate increases, it then plateaus until the next threshold at about 65\%. For pine and oak it is not possible to estimate a threshold, as there are no extreme changes in the positive part of the slope of the log hazard rate. Under scenario 2
we could only detect a threshold for spruce with a 95\% Bayesian confidence interval between 40 and 83\% (Figure~\ref{fig:loghazard}). 
 
\section{Discussion} 
We have been able to partially answer our two main questions.
Models with the defoliation effect added to selected environmental predictors (model~\ref{mod:fulldefol})  
{\em always} perform best in terms of AIC and deviance explained. Hence the environmental predictors don't entirely replace defoliation.  Environmental drivers differ by species, but some are selected for most species and scenarios (Table~\ref{Tab:selvar}). Minimum temperature in May is selected for all species except oak, and aridity index is selected  for spruce, fir and beech.   
The patterns we see in Figure~\ref{fig:loghazard} are not really consistent with defoliation mitigating the effects of climate stress, because the hazard alway increases. Mortality was dominated by storm damage. Models including storm damage as an event of interest (scenario 1) helped in identifying meaningful defoliation thresholds associated with irreversible damage ultimately leading to tree death. Such thresholds were found for Norway spruce,  fir and beech. Excluding storm damage as event of interest (scenario 2) only permitted the estimation of such a threshold for Norway spruce.
 This implies that forest management action could be based on the threshold location. 
 
The thresholds values with confidence intervals have practical relevance for forest management. Firstly, trees with defoliation greater than this threshold value but not dead yet could be harvested in an anticipatory manner. This accelerated salvage harvesting could help in avoiding wood depreciation which would otherwise occur. If the trees are harvested much later after their death, this may result in stem wood discolouration, fungi or stem rot. This is especially the case for Norway Spruce. Secondly, if trees are assessed to be partially defoliated below the critical threshold, attempts to stabilize them by forest management interventions are in general justified, since we know, with reasonable reliability, that they are not condemned to later mortality. Although it is questionable in general if it makes sense to remove healthy neighbour trees in order to stabilize those partially defoliated trees, thinning in this way can make sense in order to improve the resource availability of the target trees, i.e. increase their availability to water, sunlight and nutrients. However, such stabilization measures may remain limited to those partially defoliated trees which are of special value for example to maintain certain species mixtures or valuable forest structures.

The interpretation  of effects of the different environmental predictors is difficult due to the fact that a lot of these variables are highly correlated. In    Figure~\ref{fig:smoothsel} we have picked some examples of plausible effects for the model selected for spruce under scenario 1.  The effect of plant year is linear and negative, with later plant years having a lower log hazard;   the effect of the minimum temperature in May (tmin$\_$May lag 0) shows a minimum at 1 degree Celsius;  the effect of the aridity index (aridity lag 0) also shows an optimal range approximately between 70 - 150 and finally the effect of julian day of budburst in the previous year (budburst lag 1) is linear and negative and shows a reduced log hazard the later it is in the year.

Also, a spatial frailty (random effect) term is included in the minimal model in order to account  for short range spatial correlation in the response.  As many of the environmental predictor variables have a spatial trend, this may lead to the effects of the environmental variables and frailty term being confounded. This phenomenon is called spatial confounding and often occurs in environmental, ecological and epidemiological applications of spatial statistics (see e.g. \cite{augustin2007spatial,paciorek2010importance,hodges2013richly,hanks2015restricted}). 
It can be shown that for fixed effects of interest, the inclusion of the spatial random effects decreases the bias of the fixed effects~\citep{dupont2022spatial+, DupontAugustin23} compared to a model not including spatial smooths or spatial random effects.  This result confirms that for model selection, the random effect should be in the minimal model - as we have done here.

The data analysed here are non-standard time-to-event data. Not all grid locations are observed yearly, grid locations can be abandoned and new grid locations are added throughout the survey. Thus the risk set will decrease when a tree dies, but the risk set can also grow since new trees can be recruited into the study at any time. 
If a tree is recruited  we know it was alive up until that point. We do not, however, have knowledge of the trees in that area that were not observed to die (or to be censored). We then have a mix of interval censoring  of approximate 3 year intervals, left truncation and right censored data up to the
nearest year.  Our approach for estimating the Cox model deals with the left truncation. 

Our approach is an improvement on methodology which as been used in the past in the context of survival analysis with trees.  In \cite{neuner2015survival} model results suggested that the climate variables have some impact on survival, namely winter temperatures and summer rainfall. However, these variables were simplistic averages and not taking into account the changes in climate throughout the time period. Neuner et al. have performed age-dependent analysis, but do not have recorded values of the climatic variables for the whole lifetime of the tree, this will increase the amounts of missing data and these models have no spatial aspect. 
Except \cite{nothdurft2013spatio} who uses B-splines to model non-linear  effects  of covariates, in all other work with a forestry context \citep{thapa2016modeling,li2015survival,lee2011comparison} linear functions of the covariates are used. 
We use calendar year as the survival time and this is in contrast to \cite{neuner2015survival} and \cite{nothdurft2013spatio}, who used tree age as the survival time in their model fitted to Baden-W\"urttemberg tree survival data. In their model the baseline hazard is a function of age.
  This assumption is good if we have all potential explanatory variables, that is, if the model is correctly specificied,  but this is usually not the case (see \cite{Pencina2007}). When using age as the survival time, the baseline hazard would explain any age related trends, or any other characteristics that are not influenced by the environment. It would therefore miss any climatic temporal trends not recorded in the data, whereas a model with calendar time as the time variable would not. For example, we do not have the climatic variables for the initial years of older trees with a maximum of 257 years age. For a model with tree age as time, we would have to use trees that are of a certain age, similar to \cite{nothdurft2013spatio} who used trees with minimum age of 60 years. This would mean some trees would not be included in the model at all, and we would have less data. 
  For these reasons, we argue that our approach 
  is  well suited for investigating the relationship between mortality and environmental variables.

In order to deal with the different statistical challenges in the data we have used a smooth additive Cox model with random effects  and were able to answer the questions posed. The statistical challenges are the large number of correlated predictor variables; time varying variables and variables with non-linear effects and short range spatial correlation. We have been able to deal with short range correlation between individual trees within a sampling grid point by including a random effect. With more than one thousand locations, this results in some very large models.  The models are big in terms of number of observations and number of parameters and parallel computing combined with marginal discretization of covariates \citep{wood2017generalized} for the Laplace approximated maximum likelihood (LAML) estimation enabled us to fit these big models in very little time. 
The large number of correlated time varying environmental predictors with non-linear effects ruled out traditional backward selection. Our proposed `boost forward penalise backward' model selection scheme based on combining component-wise gradient boosting \citep{schmid2008boosting,mayr2012generalized} with integrated backward selection \citep{marra2011practical} is promising.
In addition, this new model selection approach also works in applications where the number of parameters is larger than the sample size. 

\section{Acknowledgement}
We thank editor, associate editor and referees for their constructive comments which helped improving the manuscript. 
\if1\blind
{
\section{Disclosure Statement} 
As part of a research collaboration between the Forest Research Institute Baden-W\"urttemberg (FRI) and the University of Bath, Nicole Augustin received funding from the FRI to carry out parts of the research presented in the paper. We don't forsee any potential conflicts arising from this. 
}  \fi

\bibliographystyle{chicago}
\bibliography{icp}

\begin{thebibliography}{}

\bibitem[\protect\citeauthoryear{Augustin, Musio, von Wilpert, Kublin, Wood,
  and Schumacher}{Augustin et~al.}{2009}]{Augetal09}
Augustin, N., M.~Musio, K.~von Wilpert, E.~Kublin, S.~Wood, and M.~Schumacher
  (2009).
\newblock Modelling spatio-temporal forest health monitoring data.
\newblock {\em Journal of the American Statistical Society\/}~{\em 104\/}(487),
  899--911.

\bibitem[\protect\citeauthoryear{Augustin, Lang, Musio, and {v}on
  Wilpert}{Augustin et~al.}{2007}]{augustin2007spatial}
Augustin, N.~H., S.~Lang, M.~Musio, and K.~{v}on Wilpert (2007).
\newblock A spatial model for the needle losses of pine-trees in the forests of
  baden-w{\"u}rttemberg: an application of bayesian structured additive
  regression.
\newblock {\em Journal of the Royal Statistical Society: Series C (Applied
  Statistics)\/}~{\em 56\/}(1), 29--50.

\bibitem[\protect\citeauthoryear{Bender, Groll, and Scheipl}{Bender
  et~al.}{2018}]{bender2018generalized}
Bender, A., A.~Groll, and F.~Scheipl (2018).
\newblock A generalized additive model approach to time-to-event analysis.
\newblock {\em Statistical Modelling\/}~{\em 18\/}(3-4), 299--321.

\bibitem[\protect\citeauthoryear{B\"uhlmann and Hothorn}{B\"uhlmann and
  Hothorn}{2007}]{BueHot2007}
B\"uhlmann, P. and T.~Hothorn (2007).
\newblock Boosting algorithms: Regularization, prediction and model fitting
  (with discussion).
\newblock {\em Statistical Science\/}~{\em 22\/}(4), 477--505.

\bibitem[\protect\citeauthoryear{Cox}{Cox}{1972}]{Cox1972}
Cox, D.~R. (1972).
\newblock Regression models and life-tables.
\newblock {\em Journal of the Royal Statistical Society. Series B
  (Methodological)\/}~{\em 34\/}(2), 187--220.

\bibitem[\protect\citeauthoryear{Damman, Herrman, K\"orver, Schr\"ock, and
  Ziegler}{Damman et~al.}{2001}]{DaHeKoe2001}
Damman, I., T.~Herrman, F.~K\"orver, H.~Schr\"ock, and C.~Ziegler (2001).
\newblock {\em Dauerbeobachtungsflächen Waldschäden im Level II-Programm -
  Methoden und Ergebnisse der Kronenansprache seit 1983}.
\newblock Bund-L\"ander-Arbeitsgruppe Level II / Arbeitskreis Krone. BMVEL,
  Bonn.

\bibitem[\protect\citeauthoryear{De~Martonne}{De~Martonne}{1926}]{de1926areism}
De~Martonne, E. (1926).
\newblock Areism and aridity index.
\newblock {\em Cr Hebd Acad Sci\/}~{\em 182}, 1395--1398.

\bibitem[\protect\citeauthoryear{de~Vries, Vel, Reinds, Deelstra, Klap,
  Leeters, Hendriks, Kerkvoorden, Landmann, Herkendell, Haussmann, and
  Erisman}{de~Vries et~al.}{2003}]{DeVries2003}
de~Vries, W., E.~Vel, G.~Reinds, H.~Deelstra, J.~Klap, E.~Leeters, C.~Hendriks,
  M.~Kerkvoorden, G.~Landmann, J.~Herkendell, T.~Haussmann, and J.~Erisman
  (2003).
\newblock {\em Intensive monitoring of forest ecosystems in Europe: 1.
  Objectives, set-up and evaluation strategy}, Volume 174(1).

\bibitem[\protect\citeauthoryear{Dietrich, Wolf, Kawohl, Wehberg, K\"andler,
  Mette, Röder, and Böhner}{Dietrich et~al.}{2017}]{DietWolKawo2017}
Dietrich, H., T.~Wolf, T.~Kawohl, J.~Wehberg, G.~K\"andler, T.~Mette,
  A.~Röder, and J.~Böhner (2017).
\newblock Temporal and spatial high-resolution climate data from 1961-2100 for
  the {G}erman {N}ational {F}orest {I}nventory ({NFI}).
\newblock {\em Annals of Forest Science\/}.

\bibitem[\protect\citeauthoryear{Dupont and Augustin}{Dupont and
  Augustin}{2023}]{DupontAugustin23}
Dupont, E. and N.~H. Augustin (2023).
\newblock Spatial confounding and spatial+ for non-linear covariate effects.
\newblock {\em Journal of Agricultural, Biological and Environmental
  Statistics\/}.

\bibitem[\protect\citeauthoryear{Dupont, Wood, and Augustin}{Dupont
  et~al.}{2022}]{dupont2022spatial+}
Dupont, E., S.~N. Wood, and N.~H. Augustin (2022).
\newblock Spatial+: a novel approach to spatial confounding (with discussion).
\newblock {\em Biometrics\/}, 1279--1290.

\bibitem[\protect\citeauthoryear{Eichhorn, Roskams, Poto\`ci\`c, Timmermann,
  Ferretti, Mues, Szepesi, Durrant, Seletkovi\'c, H-W.Schr\"ock, Nevalainen,
  Bussotti, Garcia, and Wulff}{Eichhorn et~al.}{2017}]{Eich2016}
Eichhorn, J., P.~Roskams, N.~Poto\`ci\`c, V.~Timmermann, Ferretti, V.~Mues,
  A.~Szepesi, D.~Durrant, I.~Seletkovi\'c, H-W.Schr\"ock, S.~Nevalainen,
  F.~Bussotti, P.~Garcia, and S.~Wulff (Eds.) (2017).
\newblock {\em ICP Forests manual on methods and criteria for harmonized
  sampling, assessment, monitoring and analysis of the effects of air pollution
  on forests.}
\newblock Th\"unen Institute of Forest Ecosystems, Eberswalde,Germany.

\bibitem[\protect\citeauthoryear{Eickenscheidt, Augustin, and
  Wellbrock}{Eickenscheidt et~al.}{2019}]{eickenscheidt2018spatio}
Eickenscheidt, N., N.~H. Augustin, and N.~Wellbrock (2019).
\newblock Spatio-temporal modelling of forest monitoring data: Modelling german
  tree defoliation data collected between 1989 and 2015 for trend estimation
  and survey grid examination using gamms.
\newblock {\em iForest Biogeosciences and Forestry\/}~{\em 12}, 338--348.

\bibitem[\protect\citeauthoryear{Federer, V\"or\"osmarty, and Fekete}{Federer
  et~al.}{2003}]{FedVor2003}
Federer, C., C.~V\"or\"osmarty, and B.~Fekete (2003).
\newblock Sensitivity of annual evaporation to soil and root properties in two
  models of contrasting complexity.
\newblock {\em Journal of Hydrometeorology\/}~{\em 4\/}(6), 1276--1290.

\bibitem[\protect\citeauthoryear{Greven and Kneib}{Greven and
  Kneib}{2010}]{greven2010behaviour}
Greven, S. and T.~Kneib (2010).
\newblock On the behaviour of marginal and conditional aic in linear mixed
  models.
\newblock {\em Biometrika\/}~{\em 97\/}(4), 773--789.

\bibitem[\protect\citeauthoryear{Guo}{Guo}{1993}]{guo1993event}
Guo, G. (1993).
\newblock Event-history analysis for left-truncated data.
\newblock {\em Sociological Methodology\/}, 217--243.

\bibitem[\protect\citeauthoryear{Hammel and Kennel}{Hammel and
  Kennel}{2001}]{HamKen2001}
Hammel, K. and M.~Kennel (2001).
\newblock {\em Forstliche Forschungsberichte München}, Chapter
  Charakterisierung und Analyse der Wasserverfügbarkeit und des
  Wasserhaushalts von Waldstandorten in Bayern mit dem Simulationsmodell
  BROOK90, pp.\  135 pp.
\newblock Technische Uni München Wissenschaftszentrum Weihenstephan, Munich,
  Germany.

\bibitem[\protect\citeauthoryear{Hanks, Schliep, Hooten, and Hoeting}{Hanks
  et~al.}{2015}]{hanks2015restricted}
Hanks, E.~M., E.~M. Schliep, M.~B. Hooten, and J.~A. Hoeting (2015).
\newblock Restricted spatial regression in practice: geostatistical models,
  confounding, and robustness under model misspecification.
\newblock {\em Environmetrics\/}~{\em 26\/}(4), 243--254.

\bibitem[\protect\citeauthoryear{Hastie and Tibshirani}{Hastie and
  Tibshirani}{1987}]{hastie1987generalized}
Hastie, T. and R.~Tibshirani (1987).
\newblock Generalized additive models: some applications.
\newblock {\em Journal of the American Statistical Association\/}~{\em
  82\/}(398), 371--386.

\bibitem[\protect\citeauthoryear{Hodges}{Hodges}{2013}]{hodges2013richly}
Hodges, J.~S. (2013).
\newblock {\em Richly parameterized linear models: additive, time series, and
  spatial models using random effects}.
\newblock CRC Press.

\bibitem[\protect\citeauthoryear{Hofner, Mayr, Robinzonov, and Schmid}{Hofner
  et~al.}{2014}]{HofMayNik2014}
Hofner, B., A.~Mayr, N.~Robinzonov, and M.~Schmid (2014).
\newblock Model-based boosting in {R}: A hands-on tutorial using the {R}
  package mboost.
\newblock {\em Computational Statistics\/}~{\em 29}, 3--35.

\bibitem[\protect\citeauthoryear{Hothorn, B\"uhlmann, Kneib, Schmid, and
  Hofner}{Hothorn et~al.}{2010}]{HotBueKne2010}
Hothorn, T., P.~B\"uhlmann, T.~Kneib, M.~Schmid, and B.~Hofner (2010).
\newblock Model-based boosting 2.0.
\newblock {\em Journal of Machine Learning Research\/}~{\em 11}, 2109--2113.

\bibitem[\protect\citeauthoryear{Kimeldorf and Wahba}{Kimeldorf and
  Wahba}{1971}]{kimeldorf1971some}
Kimeldorf, G. and G.~Wahba (1971).
\newblock Some results on tchebycheffian spline functions.
\newblock {\em Journal of Mathematical Analysis and Applications\/}~{\em
  33\/}(1), 82--95.

\bibitem[\protect\citeauthoryear{Lee, Zeng, Thompson, and Dean}{Lee
  et~al.}{2011}]{lee2011comparison}
Lee, T.~C., L.~Zeng, D.~J. Thompson, and C.~Dean (2011).
\newblock Comparison of imputation methods for interval censored time-to-event
  data in joint modelling of tree growth and mortality.
\newblock {\em Canadian Journal of Statistics\/}~{\em 39\/}(3), 438--457.

\bibitem[\protect\citeauthoryear{Li, Hong, Thapa, and Burkhart}{Li
  et~al.}{2015}]{li2015survival}
Li, J., Y.~Hong, R.~Thapa, and H.~E. Burkhart (2015).
\newblock Survival analysis of loblolly pine trees with spatially correlated
  random effects.
\newblock {\em Journal of the American Statistical Association\/}~{\em
  110\/}(510), 486--502.

\bibitem[\protect\citeauthoryear{Maringer, Stelzer, Paul, and
  Albrecht}{Maringer et~al.}{2021}]{maringer2021ninety}
Maringer, J., A.-S. Stelzer, C.~Paul, and A.~T. Albrecht (2021).
\newblock Ninety-five years of observed disturbance-based tree mortality
  modeled with climate-sensitive accelerated failure time models.
\newblock {\em European Journal of Forest Research\/}~{\em 140\/}(1), 255--272.

\bibitem[\protect\citeauthoryear{Marra and Wood}{Marra and
  Wood}{2011}]{marra2011practical}
Marra, G. and S.~N. Wood (2011).
\newblock Practical variable selection for generalized additive models.
\newblock {\em Computational Statistics \& Data Analysis\/}~{\em 55\/}(7),
  2372--2387.

\bibitem[\protect\citeauthoryear{Mayr, Fenske, Hofner, Kneib, and Schmid}{Mayr
  et~al.}{2012}]{mayr2012generalized}
Mayr, A., N.~Fenske, B.~Hofner, T.~Kneib, and M.~Schmid (2012).
\newblock Generalized additive models for location, scale and shape for high
  dimensional data—a flexible approach based on boosting.
\newblock {\em Journal of the Royal Statistical Society: Series C (Applied
  Statistics)\/}~{\em 61\/}(3), 403--427.

\bibitem[\protect\citeauthoryear{Neuner, Albrecht, Cullmann, Engels, Griess,
  Hahn, Hanewinkel, H{\"a}rtl, K{\"o}lling, Staupendahl, et~al.}{Neuner
  et~al.}{2015}]{neuner2015survival}
Neuner, S., A.~Albrecht, D.~Cullmann, F.~Engels, V.~C. Griess, W.~A. Hahn,
  M.~Hanewinkel, F.~H{\"a}rtl, C.~K{\"o}lling, K.~Staupendahl, et~al. (2015).
\newblock Survival of norway spruce remains higher in mixed stands under a
  dryer and warmer climate.
\newblock {\em Global Change Biology\/}~{\em 21\/}(2), 935--946.

\bibitem[\protect\citeauthoryear{Nothdurft}{Nothdurft}{2013}]{nothdurft2013spatio}
Nothdurft, A. (2013).
\newblock Spatio-temporal prediction of tree mortality based on long-term
  sample plots, climate change scenarios and parametric frailty modeling.
\newblock {\em Forest Ecology and Management\/}~{\em 291}, 43--54.

\bibitem[\protect\citeauthoryear{Paciorek}{Paciorek}{2010}]{paciorek2010importance}
Paciorek, C.~J. (2010).
\newblock The importance of scale for spatial-confounding bias and precision of
  spatial regression estimators.
\newblock {\em Statistical science: a review journal of the Institute of
  Mathematical Statistics\/}~{\em 25\/}(1), 107.

\bibitem[\protect\citeauthoryear{Pencina, Larson, and D’Agostino}{Pencina
  et~al.}{2007}]{Pencina2007}
Pencina, M., M.~Larson, and R.~D’Agostino (2007).
\newblock Choice of time scale and its effect on significance of predictors in
  longitudinal studies.
\newblock {\em Statistics in Medicine\/}~{\em 26}, 1343--1359.

\bibitem[\protect\citeauthoryear{Pinheiro and Bates}{Pinheiro and
  Bates}{2006}]{batespinheiro2007}
Pinheiro, J.~C. and D.~M. Bates (2006).
\newblock {\em Mixed-effects models in S and S-PLUS}.
\newblock Springer science \& business media.

\bibitem[\protect\citeauthoryear{Schmid and Hothorn}{Schmid and
  Hothorn}{2008}]{schmid2008boosting}
Schmid, M. and T.~Hothorn (2008).
\newblock Boosting additive models using component-wise p-splines.
\newblock {\em Computational Statistics \& Data Analysis\/}~{\em 53\/}(2),
  298--311.

\bibitem[\protect\citeauthoryear{Silverman}{Silverman}{1985}]{Silv85}
Silverman, B. (1985).
\newblock Some aspects of the spline smoothing approach to nonparametric
  regression curve fitting.
\newblock {\em Journal of the Royal Statistical Society, Series B\/}~{\em 47},
  1--52.

\bibitem[\protect\citeauthoryear{Thapa, Burkhart, Li, and Hong}{Thapa
  et~al.}{2016}]{thapa2016modeling}
Thapa, R., H.~E. Burkhart, J.~Li, and Y.~Hong (2016).
\newblock Modeling clustered survival times of loblolly pine with
  time-dependent covariates and shared frailties.
\newblock {\em Journal of Agricultural, Biological, and Environmental
  Statistics\/}~{\em 21\/}(1), 92--110.

\bibitem[\protect\citeauthoryear{Thi\'ebaut and B\'enichou}{Thi\'ebaut and
  B\'enichou}{2004}]{Thiebaut2004}
Thi\'ebaut, A.~C. and J.~B\'enichou (2004).
\newblock Choice of time-scale in cox's model analysis of epidemiologic cohort
  data: a simulation study.
\newblock {\em Statistics in Medicine\/}~{\em 23}, 3803--3820.

\bibitem[\protect\citeauthoryear{Tian, Zucker, and Wei}{Tian
  et~al.}{2005}]{tian2005cox}
Tian, L., D.~Zucker, and L.~Wei (2005).
\newblock On the cox model with time-varying regression coefficients.
\newblock {\em Journal of the American statistical Association\/}~{\em
  100\/}(469), 172--183.

\bibitem[\protect\citeauthoryear{Whitehead}{Whitehead}{1980}]{whitehead1980fitting}
Whitehead, J. (1980).
\newblock Fitting cox's regression model to survival data using glim.
\newblock {\em Applied Statistics\/}, 268--275.

\bibitem[\protect\citeauthoryear{Wood}{Wood}{2017}]{Wood17}
Wood, S. (2017).
\newblock {\em Generalized Additive Models. An Introduction with R. Second
  Edition.}
\newblock Chapman \& Hall/CRC, Boca Raton.

\bibitem[\protect\citeauthoryear{Wood}{Wood}{2011}]{wood2011fast}
Wood, S.~N. (2011).
\newblock Fast stable restricted maximum likelihood and marginal likelihood
  estimation of semiparametric generalized linear models.
\newblock {\em Journal of the Royal Statistical Society: Series B (Statistical
  Methodology)\/}~{\em 73\/}(1), 3--36.

\bibitem[\protect\citeauthoryear{Wood}{Wood}{2013a}]{wood2012p}
Wood, S.~N. (2013a).
\newblock On p-values for smooth components of an extended generalized additive
  model.
\newblock {\em Biometrika\/}~{\em 100\/}(1), 221--228.

\bibitem[\protect\citeauthoryear{Wood}{Wood}{2013b}]{wood2013simple}
Wood, S.~N. (2013b).
\newblock A simple test for random effects in regression models.
\newblock {\em Biometrika\/}~{\em 100\/}(4), 1005--1010.

\bibitem[\protect\citeauthoryear{Wood, Li, Shaddick, and Augustin}{Wood
  et~al.}{2017}]{wood2017generalized}
Wood, S.~N., Z.~Li, G.~Shaddick, and N.~H. Augustin (2017).
\newblock Generalized additive models for gigadata: modeling the uk black smoke
  network daily data.
\newblock {\em Journal of the American Statistical Association\/}~{\em
  112\/}(519), 1199--1210.

\bibitem[\protect\citeauthoryear{Wood, Pya, and S{\"a}fken}{Wood
  et~al.}{2016}]{wood2016smoothing}
Wood, S.~N., N.~Pya, and B.~S{\"a}fken (2016).
\newblock Smoothing parameter and model selection for general smooth models.
\newblock {\em Journal of the American Statistical Association\/}~{\em
  111\/}(516), 1548--1563.

\bibitem[\protect\citeauthoryear{Zhou and Hanson}{Zhou and
  Hanson}{2018}]{ZhouHanson2018}
Zhou, H. and T.~Hanson (2018).
\newblock A unified framework for fitting bayesian semiparametric models to
  arbitrarily censored survival data, including spatially referenced data.
\newblock {\em Journal of the American Statistical Association\/}~{\em
  113\/}(522), 571--581.

\end{thebibliography}

\end{document}